\documentclass[pre,twocolumn,superscriptaddress,floatfix,notitlepage,longbibliography]{revtex4-1}
\pdfoutput=1

\usepackage[utf8]{inputenc}
\usepackage[pdftex]{color,graphicx,hyperref}
\usepackage{longtable}
\usepackage{amsmath}

\newcommand{\eref}[1]{Eq.~(\ref{#1})}

\newcommand{\fref}[1]{Fig.~\ref{#1}}
\newcommand{\tref}[1]{Table~\ref{#1}}

\begin{document}
\title{Cumulative effects of triadic closure and homophily in social networks}

\author{Aili Asikainen}
\affiliation{Department of Computer Science, School of Science, Aalto University, FI-00076, Finland}
\author{Gerardo I\~niguez}
\affiliation{Department of Computer Science, School of Science, Aalto University, FI-00076, Finland}
\affiliation{Department of Network and Data Science, Central European University, H-1051 Budapest, Hungary}
\affiliation{Next Games, FI-00530 Helsinki, Finland}
\affiliation{Instituto de Investigaciones en Matem{\'a}ticas Aplicadas y en Sistemas, Universidad Nacional Aut{\'o}noma de  M{\'e}xico, CDMX-01000, Mexico}
\author{Kimmo Kaski}
\affiliation{Department of Computer Science, School of Science, Aalto University, FI-00076, Finland}
\affiliation{The Alan Turing Institute, British Library, London NW1 2DB, UK}
\author{Mikko Kivel\"a}
\affiliation{Department of Computer Science, School of Science, Aalto University, FI-00076, Finland}

\date{\today}

\begin{abstract}
Much of the structure in social networks has been explained by two seemingly independent network evolution mechanisms: triadic closure and homophily. While it is common to consider these mechanisms separately or in the frame of a static model, empirical studies suggest that their dynamic interplay is the very process responsible for the homophilous patterns of association seen in off- and online social networks. By combining these two mechanisms in a minimal solvable dynamic model, we confirm theoretically the long-held and empirically established hypothesis that homophily can be amplified by the triadic closure mechanism. This research approach allows us to estimate how much of the observed homophily in various friendship and communication networks is due to amplification for a given amount of triadic closure. We find that the cumulative advantage-like process leading to homophily amplification can, under certain circumstances, also lead to the widely documented core-periphery structure of social networks, as well as to the emergence of memory of previous homophilic constraints (equivalent to hysteresis phenomena in physics). The theoretical understanding provided by our results highlights the importance of early intervention in managing at the societal level the most adverse effects of homophilic decision-making, such as inequality, segregation and online echo chambers.
\end{abstract}

\maketitle

One of the most important traits of human sociality is homophily~\cite{Mcpherson2001Birds}, the tendency of similar people to be connected to each other due to their shared biological and cultural attributes such as gender, occupation, or political affiliation. It has been observed across various social networks~\cite{Aukett1988Gender,Shrum1988Friendship,Elkins1993Gender,Reeder2003Effect,Thelwall2008Social,Lewis2008Tastes,Volkovich2014Gender} and it is a major force behind several pressing social issues including inequality, segregation, and online echo chambers~\cite{Zeltzer2015Gender,Vicario2016Echo,Schmidt2017Anatomy}. Thus a thorough quantitative understanding of the network mechanisms leading to homophily~\cite{Dimaggio2012Network,Diprete2006Cumulative,Currarini2009Economic,Currarini2016Simple} is essential for promoting a sufficient flow of information~\cite{Karimi2018Homophily} and equal opportunity in social networks of individuals with diverse personal preferences.

The homophily observed in social networks is often attributed either to \emph{choice homophily}, defined by people's preference when choosing whom to connect with, or to \emph{induced homophily}, rising from the constraints in the opportunities of individuals to make connections~\cite{Mcpherson1987Homophily}. However, as suggested by longitudinal empirical results \cite{Kossinets2006Empirical,Kossinets2009Origins}, these two mechanisms of homophily generation cannot be separated without considering the cumulative advantage-like dynamics~\cite{Diprete2006Cumulative} driving the evolution of social networks: Choice homophily creates circumstances for induced homophily, such as groups of similar people, which are then further reinforced in cycles of choice and induced homophily. 
While the dynamics of homophily is well understood in the case of \textit{tipping point} models of residential segregation~\cite{Schelling1971Dynamic,Vinkovic2006Physical}, a similar understanding of the dynamics in social networks is still needed in order to validate and measure homophily amplification.

\begin{figure}[t]
\includegraphics[width = 0.48 \textwidth]{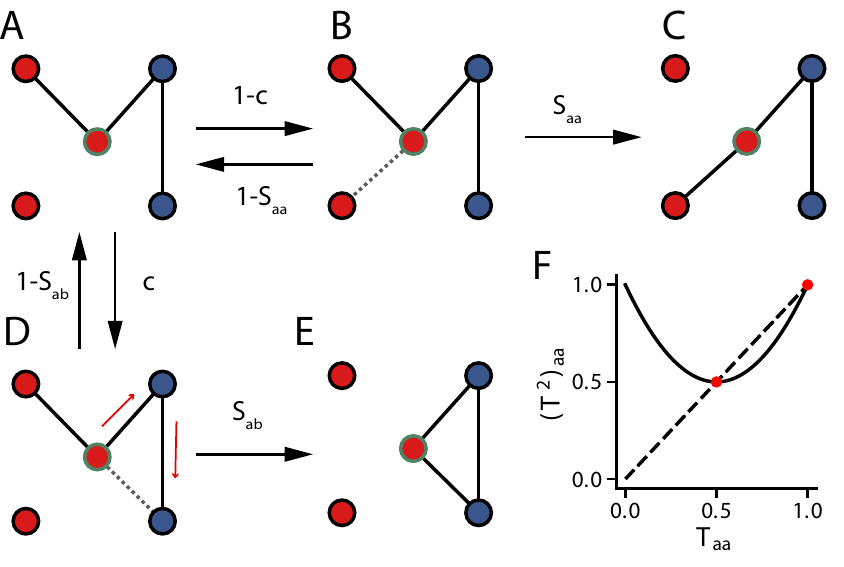}
\caption{ %
{\bf Mechanisms of triadic closure and choice homophily.}
{\bf (A)} A focal node selected uniformly at random (green boundary) finds a candidate neighbour by either {\bf (B)} selecting a node uniformly at random with probability $1 - c$, or by {\bf (D)} closing a triangle with probability $c$. {\bf (C, E)} For a focal node in group $i$, the candidate neighbour in group $j$ is accepted with probability $S_{ij}$ (where $i, j \in \{a, b\}$). We parametrise $S_{ij}$ with tunable parameters $s_a, s_b$ such that $S_{aa} = s_a$, $S_{bb} = s_b$, $S_{ab} = 1 - s_a$, and $S_{ba} = 1 - s_b$. If the potential edge (dashed line) is accepted, an edge of the focal node (selected uniformly at random) is replaced by one between the focal and candidate neighbour nodes. Otherwise no edges are rewired. 
{\bf (F)} Probability $(T^2)_{aa}$ of choosing a candidate neighbour with triadic closure from the same group as the focal node as a function of observed homophily $T_{aa}$ for equally sized ($n_a = n_b$) and equally connected groups ($T_{aa} = T_{bb}$ and $T_{ab} = T_{ba} = 1 - T_{aa}$). 
If the network is not randomly mixed ($T_{aa} \neq 1/2$), the probability of triadic closure choosing two nodes of the same group is always larger than the same probability if the selection is done uniformly at random (1/2), implying that triadic closure amplifies existing homophily in the network.  
However, triadic closure without choice homophily is not enough to maintain the observed homophily in the network, which would make $(T^2)_{aa}$ equal to the observed homophily (dotted and solid lines cross). An exception is the case $T_{aa}=1$, where two completely separate components exist and triadic closure cannot create edges between them.
}
\label{fig:schematic}
\end{figure}

Here we introduce a minimal model of social network evolution to analyse to what extent the structural constraints caused by triadic closure affect observed homophily. The triadic closure mechanism has been reported as the most common structural constraint \cite{Kossinets2006Empirical} and can explain many salient features of empirical social networks. These include a high number of closed triangles between acquaintances and fat-tailed degree distributions~\cite{Toivonen2009Comparative,Gong2012Evolution,Klimek2013Triadic,Bianconi2014Triadic}. Thus triadic closure should be considered as the main mechanism in most minimal dynamic social network models~\cite{Granovetter1973Strength,Toivonen2009Comparative,Snijders2011Statistical}. In our approach, individuals are considered to belong to either of the two groups representing the values of a static attribute of interest (gender, class, party, etc.) and they are let to rewire their connections by two mechanisms: triadic closure (modelling the creation of edges via current contacts) and random rewiring [emulating any unknown mechanisms beyond triadic closure, such as focal closure in large foci \cite{Kossinets2006Empirical,Kumpula2007Emergence}]. Choice homophily/heterophily is implemented by accepting new links with a bias probability dependent on the similarity of attributes between individuals (see \fref{fig:schematic} and Materials and Methods [MM] for a more detailed model definition and parameters).

In this study we characterise the rich tapestry of emergent behaviour captured by the model with a mean-field bifurcation analysis for varying relative group sizes, triadic closure probabilities, and choice homophily rates. By tuning the parameters of the system with empirical data on friendship and communication networks, we find that, under the right circumstances, even a small amount of choice homophily may be greatly amplified by triadic closure. This suggests that the observations of homophilous patterns of association in society should not be interpreted solely on the basis of a human preference for similarity, but as a constantly evolving interplay between structural constraints and homophily, one that requires computational simulation as a central part of the analysis.

\section*{Results}

\begin{figure*}[t]
\includegraphics[width = \textwidth]{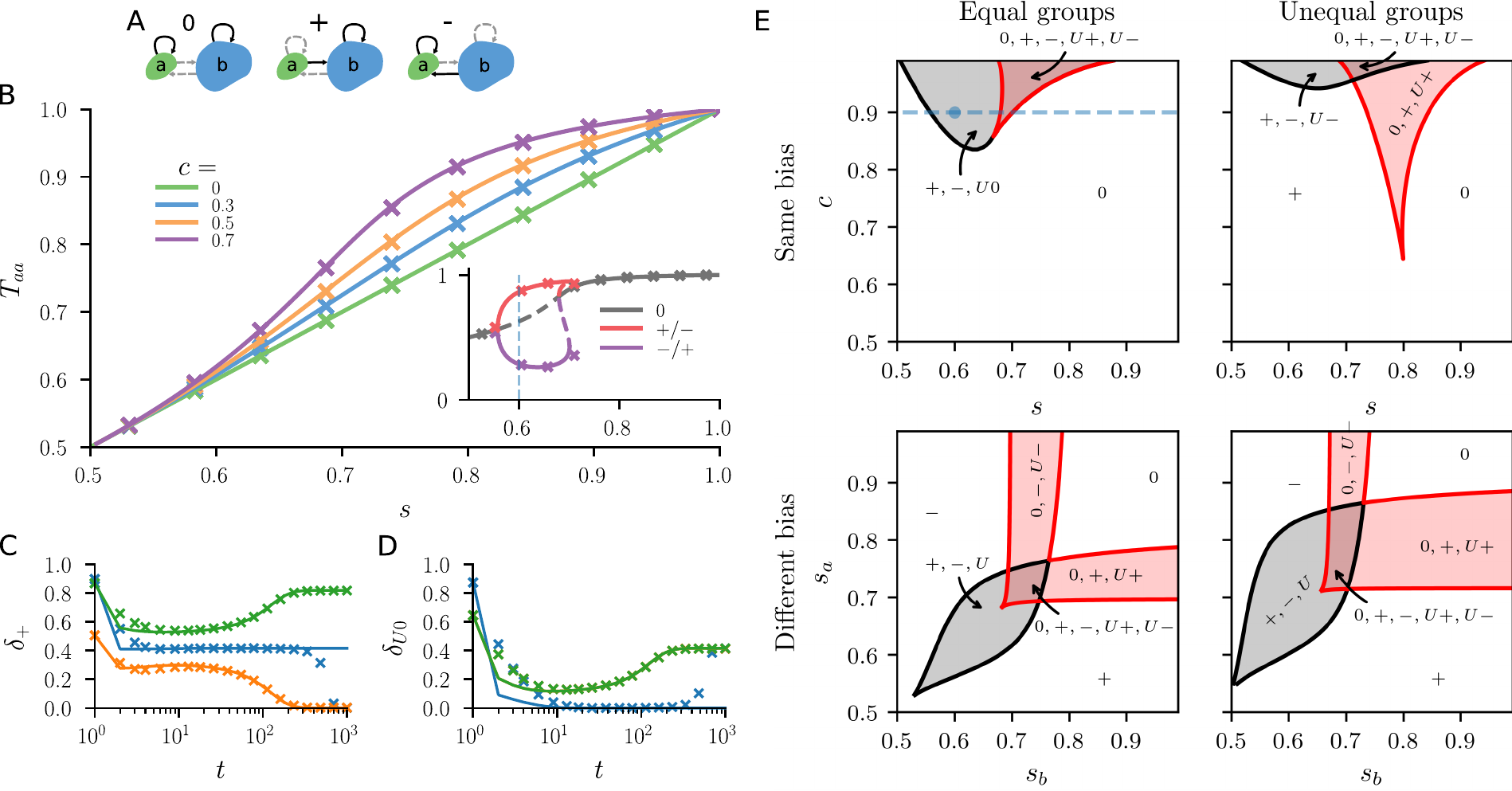}
\caption{{\bf Interplay between triadic closure and homophily.}
{\bf (A)} Stable fixed points of the model based on bifurcation analysis (see SI) and classified by large/low intra- and inter-group connectivity (continuous/dashed lines) relative to other fixed points ($n_b > n_a$). Fixed point $0$ has many edges within groups and few between, while in fixed point $+$ ($-$) the large (small) group forms the core and attracts edges from the small (large) periphery group. These fixed points can often be characterised as {\it homophily amplification} or {\it core-periphery structure}. Stable points have saddle fixed points as unstable variants ($U0$, $U+$, and $U-$) with qualitatively similar relative connectivities.
{\bf (B)} Fraction of neighbours $T_{aa}$ in the same group of a focal node in group $a$, as a function of choice homophily $s = s_a = s_b$ for $n_a = n_b$  and varying triadic closure $c$. Mean-field calculations (lines) and numerical simulations (crosses) agree perfectly (see SI for more systematic analysis). For these values of $c$ the only stable point is $0$, such that increasing triadic closure leads to higher homophily amplification. Inset: Case $c = 0.9$ where both stable/unstable points (continuous/dashed lines) exist. Since $n_a = n_b$ fixed points $+/-$ are equivalent, but for suitable choice homophily $0.56 < s < 0.70$ one group becomes the core and the other periphery.
{\bf (C-D)} Euclidean distance $\delta$ in $(T_{aa}, T_{bb})$-space between current state of the dynamics and fixed points $+$ (C) and $U0$ (D) as a function of time $t$ (in units of the average number of times an edge is selected for rewiring) for $n_a = n_b$, $c = 0.9$, $s = 0.6$ (dashed vertical line in panel B inset), and several initial network configurations. The dynamics amplifies homophily first by quickly approaching the unstable point $U0$, and then by slowly approaching a core-periphery stable point ($+/-$). For one of the initial conditions chosen, the dynamics approaches $-$ rather than $+$.
{\bf (E)} Phase diagrams of available fixed points in $(s, c)$-space (upper row, $s = s_a = s_b$) and $(s_b, s_a)$-space (lower row, $c = 0.95$), both for $n_a = n_b$ (left column) and $n_a = 0.1$ (right column). Dashed line corresponds to parameter values in panel B inset, and the dot to panels C and D. Panel B corresponds to the region with a single fixed point (0) in the upper left diagram.
}
\label{fig:mf}
\end{figure*}

In our approach, the interplay between homophily and triadic closure in a social network with two attribute groups forms a dynamical system, in which the evolution from an arbitrary initial network depends on the parameters that regulate choice homophily and triadic closure. As with any other network model, the associated stochastic process exhibits random fluctuations, but the average dynamics of key behavioural quantities such as the degree distribution, clustering coefficient, and centrally, the observed homophily, depend deterministically on the parameters. We characterise the behaviour of the system by using a mean-field bifurcation analysis and confirm our results with numerical simulations (see MM and Supplementary Information [SI]).

When individuals form new edges in a network with existing homophilous patterns of connectivity, they link to their own group even without being explicitly homophilic (\fref{fig:schematic}F), a tendency that is even more prominent in the presence of a choice homophily bias. This process increases the observed homophily in the network beyond that due to choice homophily, which in turn increases the likelihood for homophilic connections in upcoming triadic closure events. The result of this cumulative advantage-like cycle is an {\it amplification} of choice homophily, as long as groups are comparable in size and edges are not exclusively created by triadic closure (\fref{fig:mf}B). In the extreme case of no random rewiring 
(lack of other mechanisms of edge creation), even a moderate choice homophily bias will segregate the social network into fully disconnected groups (see SI for a derivation of this result).

In addition to homophily amplification, the triadic closure mechanism and choice homophily may also lead to a {\it core-periphery}~\cite{Borgatti2000Models,Holme2005Core,Rombach2017Core} social structure where the core group connects almost exclusively with itself, while the periphery group connects with the core group even in the presence of choice homophily (\fref{fig:mf}B inset). This effect, seemingly opposed to the drive of individuals to find homophilous connections in the periphery group, is due to the large likelihood of finding a candidate neighbour in the core group while attempting to close triangles (\fref{fig:schematic}D). A core-periphery social structure is possible when choice homophily biases are moderate and triadic closure is high, or when there is asymmetry in the social network due to unequal group sizes or choice homophily biases (Fig. \ref{fig:mf}E)

The rise of homophily amplification and core-periphery social structures depends not only on the parameters regulating triadic closure and choice homophily, but on the initial conditions and random fluctuations of network evolution, meaning that the system exhibits memory of previous structural constraints, or {\it homophily hysteresis} (see \fref{fig:mf}B for an example, and \fref{fig:mf}E for a more systematic analysis). In other words, the social network can experience dramatic non-reversible changes if the system parameters change, suggesting that real social networks may have persistent memory of homophily, with a structure dependent both on current choice homophily biases and their history. Therefore, the timing of interventions aiming to reduce observed homophily or the formation of core groups in, say, an online social network, can be critical. Once the network has reached a stable point of its dynamics or is close to one, it can be much more difficult to drive the system to another stable point by attempting to change the choice homophily of individuals or other parameters.

The time scales at which the social network is driven towards homophily amplification or a core-periphery structure vary greatly (\fref{fig:mf}C). Homophily amplification is generally a fast process, requiring only a few rewiring events per edge for the social network to reach a stable point. On the other hand, a core-periphery social structure evolves slowly towards equilibrium, and even if a network would eventually stabilise to a core-periphery structure it
may exhibit fast homophily amplification first (\fref{fig:mf}D). This result suggests that real social networks, both in society and online platforms, might not show a stable, fully realised core-periphery structure, but a transient state slowly drifting towards the structural dominance of one group over the other.
In the case that the network is first driven towards homophily amplification, the group that eventually becomes the core can depend purely on random chance (\fref{fig:mf}C).

\begin{figure}[t]
\includegraphics[width = 0.48 \textwidth]{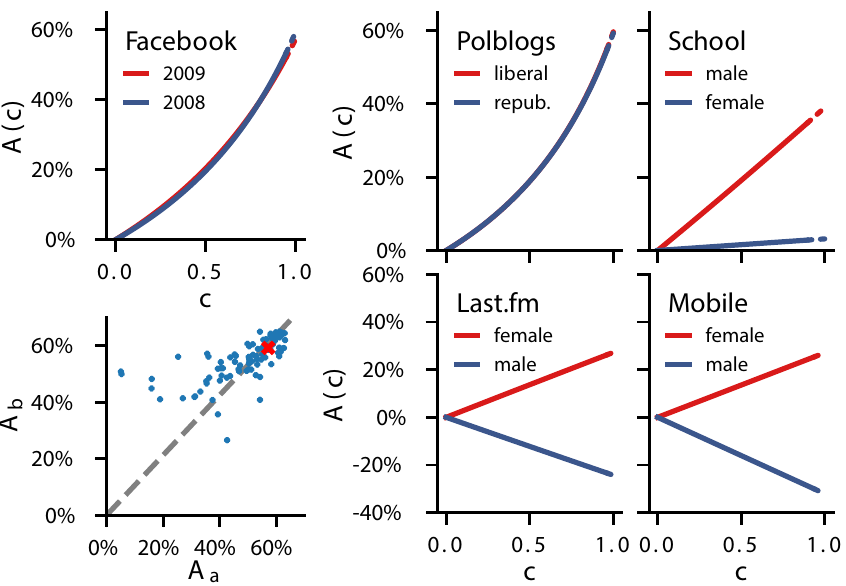}
\caption{{\bf Homophily amplification in real-world social networks.}
Estimated relative homophily amplification $A(c) = [ \hat{s}(0)-\hat{s}(c) ] / |\hat{s}(0)|$ of each attribute group in several empirical datasets, where $\hat{s}(c) = 2s(c) - 1$ is the estimate of choice homophily $s$ for given triadic closure $c$, normalized such that $\hat{s}=0$ implies no homophily/heterophily in the network. Edges, nodes and attributes in each dataset are: Facebook friendships for users by graduation year in the largest of the 100 networks (Facebook), links between political blogs by party (Ploblogs), friendships between students by gender (School), Last.fm friendships for users by gender (Last.fm), and calls between individuals by gender (Mobile).
The small group $a$ is denoted by a red line. %
Scatter plot: Correlation between maximum homophily amplification $A_i(1)$ for groups $i = a, b$ in 100 universities of the Facebook dataset ($n_a < n_b$). The largest network is marked with a red cross. See MM and \tref{tab:data} for dataset details.}
\label{fig:data}
\end{figure}

In order to estimate how much the observed homophily differs from choice homophily in real-world social networks, as well as to find the stable point that best corresponds to their structure, we fit several empirical datasets of off- and online social interactions to our model of triadic closure and choice homophily (\fref{fig:data}). In all studied cases, we observe that both groups are homophilic and we estimate that they are also intrinsically homophilic (in terms of choice homophily). Three of them show homophily amplification in both groups: a Facebook friendship network consisting of two classes in a US university~\cite{Traud2012facebook100}, a one-day contact network of primary school students divided by gender~\cite{Stehle2011Sociopatterns}, and a network of political blogs divided by party affiliation~\cite{Adamic2005polblogs}. The rest of them -- a friendship network in a website for sharing music listening habits (Last.fm) and a mobile phone call network, both divided by gender~\cite{Onnela2007Structure} -- display a pattern where part of the observed homophily within the smaller female group could be explained by homophily amplification, but the choice homophily in the larger male population could be underestimated due to the triadic closure mechanism.

The maximum relative difference we measure between the estimated choice homophily and the observed homophily, $A(1)$, goes up to around $60\%$ for the social network of political blogs and the largest Facebook network, with each exact estimate of choice homophily depending on the latent tendency for triadic closure in the system ($c$).  
However, the parity of the amplification [$A(c)$] is independent of this estimate, and the growth of amplification is monotonous as a function of $c$. A systematic analysis of 100 Facebook networks \cite{Traud2012facebook100} reveals that the maximum estimated values for amplification [$A(1)$] are all positive and that the largest network is a typical example among these networks (\fref{fig:data}). 
Our results suggest that using observed homophily as a naive estimator for choice homophily can lead to a significant overestimation or underestimation of the intensity of homophily (even for a moderate amount of triadic closure) in several real-world social networks, both in society and online platforms.

\section*{Discussion}

Our findings show that the homophilous patterns of association typically seen in empirical social networks arise not only due to an individual preference for similarity, but are the result of a cumulative advantage-like process that has the tendency to amplify this intrinsic bias for choice homophily due to triadic closure. By means of a minimal model of social network evolution, we find bounds of the amounts of triadic closure and choice homophily necessary for such amplification of homophily to arise. This corroborates theoretically previous observations in organisational~\cite{Mcpherson1991Evolution} and communication~\cite{Kossinets2006Empirical,Kossinets2009Origins} networks. In the generic case of a moderate amount of triadic closure events and similarly sized attribute groups, choice homophily is amplified by triadic closure through a tipping point mechanism analogous to the one responsible for residential segregation in the Schelling model~\cite{Schelling1971Dynamic}, in which segregation takes place in the social network topology rather than in the physical space.

In addition to the %
homophily amplification, our results suggest that the interplay between triadic closure and choice homophily is a plausible explanation for the emergence of %
the core-periphery structure found in social, communication, academic, trade, and financial networks~\cite{Borgatti2000Models,Holme2005Core,Rombach2017Core}. In such structures, the core group of individuals is so well connected that following the edges via triadic closure almost always leads to the same group, making the core even more connected. While triadic closure and homophily are already considered as the contributing factors in %
the formation of communities (cohesive and assortative groups densely connected within), the impact of node attributes on the core-periphery structure is %
less studied. Our model implies that the dynamic transition to core-periphery networks is slow and often preceded by the fast but temporary homophily amplification. This %
may partly explain why the social networks literature has focused on clustered networks rather than other, %
rarer types of intermediate-scale structures.

The coupled effects of triadic closure and choice homophily we observe, include also the memory of homophilic constraints,  i.e. systems with multiple and co-existing stable points for a wide range of relative group sizes and amounts of triadic closure and choice homophily. In other words, social networks may preserve memory of their previous structural configurations, making it difficult to alter the shape of a stable network, %
for example, by varying the typical choice homophily of its individuals. Our findings suggest that, when planning external intervention in order to reduce homophily-induced social segregation, the action measures should be taken sooner rather than later, since %
the scale of interventions with meaningful effect on the structure of the social network increases with time.

As the choice homophily is not directly observable from a static social network data, one needs to infer it from the available information. Such inference is always subject to assuming a model for the data creation, and the exact estimates for the choice homophily should always be interpreted with this in mind. For example, our stylized %
model and estimation procedure assumes that the network is in a stable state and that there are no other mechanisms affecting the system. 
Furthermore, our fitting is simply based on matching the linking probabilities $T$ of the data with %
the stable states of the model. More elaborate fitting could be done (and has been done for similar models \cite{Chen2018Flexible}) using approximate Bayesian computation methods \cite{Lintusaari2017Fundamentals}, which would allow one to include more observables to be matched. It would also allow fitting for more parameters, such as the time since the start of the model $t$ or ones needed for additional mechanisms.

In contrast %
to our approach, stationary, non-mechanistic models, such as exponential random graphs (ERGMs), can also be used to study the interplay between %
triadic closure and homophily \cite{Foster2011Clustering,Peng2015Assortative,Mele2017Structural2}.  The key conceptual difference to our approach is that ERGMs are static network models, which can be used to balance between the tendency towards triangles, homophilic edges, homophilic trianges, and many other network features as factors explaining the network structure, whereas our approach is a model of cumulative interplay of two explicit and microscopic network evolution mechanisms. (Note, however, that some carefully crafted microscopic network evolution models can converge, under certain assumptions, to ERGMs as stable states \cite{Mele2017Structural1,Mele2017Structural2}.) This means that ERGMs do	not explicitly model cumulative processes or tell anything about multiple time scales or meta-stable states, which we find as a consequence of combining triadic closure with the choice homophily bias. 
 Furthermore, the hysteresis phenomenon, i.e. the memory effects and multiple qualitatively different stable states for the same paramater values, we find emerging naturally as a consequence of triadic closure and homophily, would be considered as an undesirable feature of an ERGM \cite{Snijders2006New}. 
In the present approach it is assumed %
that the biological and cultural attributes underlying the homophily are constant in time. While this assumption is almost true for long-term individual characteristics such as gender or religion, it is less so for traits like political affiliation, occupation, and opinions. Networks where both the edges and attributes change adaptively to each other, i.e. following %
adaptive co-evolutionary dynamics, have been studied extensively for biological, economic, and social phenomena~\cite{Gross2008Adaptive,Gross2009Adaptive,Zschaler2012Adaptive}. When the edges between nodes with similar attributes are favoured, the adaptive dynamics self-organise into heterogeneous networks where groups of individuals %
sharing attributes are structurally distinguishable from each other~\cite{Ito2001Spontaneous,Bornholdt2003Self,Holme2006Nonequilibrium,Vazquez2007Time,Iniguez2009Opinion}. Such a generic feature of adaptive networks makes it likely that our observations of the cumulative effects of triadic closure and homophily will hold even in the case of time-dependent individual attributes. Still, the study of an adaptive interplay between triadic closure and homophily is a worthy line of future research that may reveal additional, complex feedback loops between social structure and attribute evolution.

The simplicity of our framework suggests that the presence of triadic closure and choice homophily for a given attribute value is enough to explain some salient features of empirical social networks like the homophily amplification and core-periphery structures. Yet, the effects of more realistic features of society %
such as the existence of more than two values for a single attribute, structural constraints beyond triadic closure, and the coexistence of several attributes in a population [in the spirit of the Axelrod model of cultural dissemination~\cite{Axelrod1997Dissemination}],remain to be studied. We anticipate that our results %
promote even more interest in the data-driven computational simulation of social interactions and shed further light on the relationship between triadic closure and homophily. This insight will help researchers and policy-makers in devising %
intervention strategies to decrease the most adverse effects of homophilic decision-making, including segregated social structures like gender-specific workplaces and partisan political systems.

\section*{Materials and Methods}
{\small

\subsection*{Model definition and parameters}

We introduce a model of social network evolution and homophily dynamics with a simultaneous interplay of triadic closure  and choice homophily. The model is stylised such that it contains a minimal amount of assumptions on how social relationships are made and forgotten, but it is otherwise maximally random. The initial social structure is a random network with static attribute groups $a$ and $b$ (of relative size $n_a$, $n_b$ with $n_a + n_b = 1$) distributed among nodes uniformly at random and independently of the initial network structure, such that there is a fraction $P_{ab} = P_{ba}$ of edges between groups, and fractions $P_{aa}$, $P_{bb}$ within each group ($P_{ab} + P_{aa} + P_{bb} = 1$).

From its initial state the network develops by the nodes updating their connections. At each time step we select a focal node uniformly at random and a candidate neighbour, representing a social encounter that might lead to a new social relationship (\fref{fig:schematic}). The candidate neighbour is chosen by triadic closure with probability $c$, or uniformly at random otherwise [emulating any other mechanisms for edge creation beyond triadic closure \cite{Davidsen2002Emergence,Vazquez2003Growing,Marsili2004Rise,Toivonen2006Model,Bianconi2014Triadic}]. The triadic closure mechanism can be algorithmically implemented in two ways~\cite{Toivonen2009Comparative}: by connecting two neighbours of the focal node, or by connecting the focal node with the neighbour of a neighbour. Since both mechanisms lead to the same mean-field description of the network evolution in the infinite size limit~\cite{Timar2016Scale}, we choose the latter for our simulations.

After checking that a new edge between the focal node and candidate neighbour would not create a multi-edge or self-loop, the edge is created with probability $S_{ab}$ if the focal node is in group $a$ and the candidate neighbor is in group $b$. The elements $S_{ab}$ form a $2 \times 2$ bias matrix specifying the amount of choice homophily/heterophily in the social network. For simplicity we parametrise the bias matrix as $S_{aa}=s_a$, $S_{ab} = 1 - s_a$, $S_{bb} = s_b$, and $S_{ba} = 1 - s_b$, where $s_a$ ($s_b$) is the choice homophily for group $a$ ($b$). In this way, when $s_a=s_b=1/2$ all of the elements of the bias matrix are also $1/2$, i.e. there is no homophily bias. Note that multiplying the bias matrix by a constant changes the speed of network evolution, but not the fixed points of the dynamics.

Finally, as maintaining social connections requires mental capacity and time investments, creating new connections implies forgetting some of the old ones \cite{Saramaki2014Persistence}. We model this process by randomly removing an edge of the focal node after a successful edge creation. Random link removal~\cite{Marsili2004Rise} does not involve additional assumptions of social behaviour, and is a typical choice in this type of social network models [along with random node deletion~\cite{Davidsen2002Emergence,Kumpula2007Emergence}].

\begin{table*}[t]
\centering
\begin{tabular}{l cccccc|cccc}
 & group by & $N$ & $\langle k \rangle$ & $n_a$ & $\text{HI}_a$ & $\text{HI}_b$ &$ s_a(0) / s_b(0)$ & $s_a(1) / s_b (1)$ \\
\hline
Facebook & class & 14086 & 49.76 & 46.5\% & 0.451 & 0.512 & 0.734/0.748 & 0.649/0.633 \\
Polblogs & party & 1224 & 27.31 & 48.0\% & 0.813 & 0.810 & 0.909/0.902 & 0.665/0.664 \\
School & gender & 212 & 4.12 & 49.1\% & 0.308 & 0.213 & 0.656/0.605 & 0.595/0.602 \\
Last.fm & gender & 188672 & 5.49 & 31.7\% & 0.168 & 0.106 & 0.621/0.540 & 0.588/0.550 \\
Mobile & gender & 2173030 & 4.35 & 44.9\% & 0.100 & 0.057 & 0.555/0.526 & 0.540/0.535
\end{tabular}
\caption{\textbf{Properties of empirical datasets.} \textnormal{List of real-world social networks used in this study, their main properties and estimated model parameters. $N$ is the number of nodes in the network, $\langle k \rangle$ the average degree, $n_a$ the fraction of nodes in the smaller attribute group, $\text{HI}_{a} = (T_{aa}-n_a)/(1-n_a)$ and $\text{HI}_{b} = (T_{bb}-n_b)/(1-n_b)$ the Coleman homophily indices~\protect\cite{Coleman1958Relational} of the groups, $s_a(0)$ and $s_b(0)$ the estimates of the bias parameters when $c=0$, and $s_a(1)$ and $s_b(1)$ the estimates when $c=1$.}}
\label{tab:data}
\end{table*}

\subsection*{Mean-field bifurcation analysis}

We derive approximate analytical expressions for the temporal evolution of the amount of observed homophily in a social network based on %
a mean-field bifurcation analysis of our model. In the case of two attribute groups, the state of the system at time $t$ can be tracked by a $2 \times 2$ matrix $P$, where the element $P_{ab}$ is %
the probability that an edge chosen uniformly at random lies between groups $a$ and $b$. Equivalently, we may follow the dynamics of a $2 \times 2$ \textit{transition} matrix $T$, where element $T_{ab}$ is %
the probability that following a random edge from a node in group $a$ leads to a node in group $b$. The probability $M_{ab}$ that, in a single time step of the dynamics, we create an edge between the nodes in groups $a$ and $b$, respectively, is given by
\begin{equation}
\label{eq:modelMat}
M_{ab}=[c (T^2)_{ab} + (1-c) n_b ] S_{ab}.
\end{equation}
We then write a rate equation describing the change in the fraction of edges within group $a$, 
\begin{equation}
\label{eq:motion}
\frac{d P_{aa}}{d t} = n_a M_{aa} - n_a T_{aa} (M_{aa} + M_{ab}),
\end{equation}
and a similar equation for $P_{ab}$. We determine the fixed points of the dynamics and their stability through linear stability analysis, and confirm the validity of the mean-field approximation via extensive simulations (see the SI for details of the analytical solution of the model, and \fref{fig:mf} for a summary of the analytical results).

\subsection*{Numerical simulations}

We use numerical simulations to verify the accuracy of the mean-field approximation of \eref{eq:motion} (\fref{fig:mf}B). We first construct a random network with $N=10^5$ nodes and the average degree $\langle k \rangle = 50$. To create networks with different initial conditions in terms of in- and out-group edges, we choose values for the fractions of the same-group neighbours $T_{aa}$ and $T_{bb}$. For simulations in \fref{fig:mf}B the initial networks have $(T_{aa},T_{bb}) = (0.5,0.5)$. For the inset we use two initial conditions, i.e. $(T_{aa},T_{bb}) = (0.1,0.9)$ and $(T_{aa},T_{bb}) = (0.9,0.1)$. We then create two random networks so that the number of edges in each network corresponds to the desired numbers of in-group edges. Finally we place the remaining edges randomly between the two groups, so that the final network has $L = N\langle k \rangle/2$ edges.

Simulations follow the model definition described above. Between times $t$ and $t + 1$ we attempt to rewire $L$ edges, so that on average each edge in the network is chosen once. For the parameters in \fref{fig:mf}B, $t = 10^2$ is enough for geting convergence to a fixed point, while for the parameters in the inset we need $t = 10^3$. Each point in \fref{fig:mf}B is averaged over $10^2$ realizations, with the standard deviations smaller than the marker size (see SI for a more detailed analysis of model parameters).

\subsection*{Social network data}

We use several large-scale social network datasets in order to determine empirically the possible effects of triadic closure on the observed homophily. The first one is \textit{Mobile}, which is a social network based on call detail %
records between millions of subscribers in a European country~\cite{Kivela2012Multiscale}. We draw edges between the individuals if they have called each other within a one-year observation period and divide them into two groups based on the gender listed in subscription data. The second one is \textit{Facebook},which is a friendship network of two classes at the University of Pennsylvania in the US. The dataset includes friendships  and metadata for 100 universities during 2005~\cite{Traud2012facebook100}. For each university we use the subnetwork of the two largest classes. The third one is \textit{Polblogs}, which is a network of political blogs collected in 2005~\cite{Adamic2005polblogs}, with edges between two nodes if at least one of the blogs links to the other. The Blogs are split into two groups using the classification of liberal and conservative blogs provided by the original study. The fourth one is \textit{School}, which is a network between students collected by automatically sensing the proximity between individuals. The original data has a 20-second time resolution for 2 days, which we aggregate into the edges by considering two nodes connected if they have been in each others' proximity for at least 20 minutes during the observation period. Nodes are split into two groups according to gender~\cite{Stehle2011Sociopatterns}. The fifth one is \textit{Last.fm}, which is a snapshot of a self-reported friendship network in a music listening website. The network is split into two groups according to gender and it includes only the users for which this information is available (see \tref{tab:data} for a summary of dataset features and SI for more details).

\subsection*{Homophily measures and model fitting}

In order to estimate the amount of choice homophily in both groups of an empirical network, we need to solve the following %
inverse problem: Given a certain value $c$ of triadic closure, we find the choice homophily parameters $s_a$ and $s_b$ in our model, which lead to the observed edge fractions between and within groups, $P_{aa}$, $P_{bb}$ and $P_{ab}$. If we set $c=0$ we get a naive estimate of the biases that does not consider any triadic closure, but, for example, corrects for a disproportionate amount of links observed within large groups as compared to small groups even if there is no intrinsic bias. Note, however, that this feature of our estimation process leads to a different size correction than the Coleman homophily index~\cite{Coleman1958Relational}. Increasing the value from $c=0$, we see how an increasing amount of triadic closure changes the estimates of the choice homophily biases (\fref{fig:data}). We solve the inverse problem by setting $d P_{aa} / d t = 0$ in \eref{eq:motion} and solving for $s_a$ and $s_b$ given the matrix $P$, or equivalently the transition matrix $T$ (See SI for a closed-form formula).

}

\appendix

\section*{Supplementary information}
\subsection{Mean field master equation}
Here we derive the mean-field master equation for the social network model presented in the main text.

The mean-field master equation is derived in terms of the (expected) transition probability matrix $T$ giving the probability that by starting from a node in group $a$ and following a uniformly random edge we end up in a node in group $b$. For parts of the derivation it is helpful to represent this same information using the edge density matrix $P$, 
where the element $P_{ab}$ is the fraction of edges between groups $a$ and $b$. The transition matrix $T$ (in the mean field) can be written in terms of $P$ as 
\begin{equation}
T_{ab} = \frac{P'_{ab}}{\sum_b P'_{ab}},
\end{equation}
where $P' = P + Diag(P)$, and $Diag(P)$ is a matrix where non-diagonal entries of $P$ has been set to zero.

Note that if there would be more than two groups then there would be more independent elements in the transition matrix $T$ than in the $P$-matrix. (When there are $N$ groups the $T$-matrix has $N(N-1)$ independent elements because by construction the rows always sum to one. For the $P$-matrix there are $N(N-1)/2$ pairs of mirror elements, e.g. $P_{ab} = P_{ba}$, and one equation that comes from the fact that the probabilities sum to one. This leads to $N^2-N(N-1)/2-1 = N(N+1)/2-1$ independent elements in the $P$-matrix.) This means that for more than two groups the $P$-matrix is not uniquely defined by the $T$-matrix. Also note that all diagonal $P$-matrices lead to $T$-matrices 
that are just identity matrices, regardless of the proportions between the different $P$-matrix elements.

In our network evolution model the system is updated such that the neighborhood of a random node is updated in each step, and this update step is then repeated large number of times. Thus, in order to write the master equation we will first write down the equations describing the change of the system in a single step.
We define a model matrix $M$, such that the 
element $M_{ab}$ %
gives the probability that a link between groups $a$ and $b$ is created in a single step of the model given that a node from group $a$ is selected. The model matrix can be written in terms of $T$ and the model parameters as  
\begin{equation}
M_{ab}=(c (T^2)_{ab} + (1-c) n_b ) S_{ab},
\end{equation}
where $n_b$ is the fraction of nodes in group $b$. 

In a single step of the model the expected number of links between groups $a$ and $b$ increases if a link is formed between them by selecting a node from either group $a$ or group $b$:
\begin{equation}
L P^+_{ab}= \begin{cases}
n_a M_{ab} + n_b M_{ba}, &\text{when}\quad a \neq b \\
n_a M_{aa}, &\text{when}\quad a = b \,,
\end{cases}
\end{equation}
where $L$ is the number of links in the network.
If a link is formed between group $a$ and any other group we remove one link connected to the initial node and the expected number of links between groups $a$ and $b$ decreases by
\begin{equation}
L P^-_{ab} = \begin{cases}
n_aT_{ab}\sum_d M_{ad} + n_bT_{ba} \sum_d M_{bd}, & \text{when}\quad a\neq b\\
n_aT_{aa}\sum_d M_{ad}, & \text{when}\quad a=b.
\end{cases}
\end{equation} 
When we set the time unit so that in a single time unit corresponds to $L$ steps in the model, we can write
the rate equation as (see Eq. 2 in the main text)
\begin{equation}
\label{eq:rateP}
\frac{\text{d} P_{ab}}{\text{d} t} = P^+_{ab}-P^-_{ab}.
\end{equation}

In the case of two groups we are left with only two independent elements of the $P$-matrix. 
In this case we can write the $P$-matrix elements in terms of the $T$-matrix
\begin{equation}
P_{aa} = \frac{T_{aa}(1-T_{bb})}{2-T_{aa}-T_{bb}}\quad \text{and}\quad P_{bb} = \frac{T_{bb}(1-T_{aa})}{2-T_{aa}-T_{bb}},
\end{equation}
where we have used $T_{ab} = 1-T_{aa}$ and $T_{ba} = 1-T_{bb}$ and set $(T_{aa}, T_{bb}) \neq (1,1)$.

We can now write the rate equation in terms of the $T$-matrix elements by using the chain rule
\begin{equation}
\label{eq:rateT}
\frac{\text{d} T_{aa}}{\text{d} t} = \frac{\partial T_{aa}}{\partial P_{aa}}\frac{\text{d} P_{aa}}{\text{d} t}+\frac{\partial T_{aa}}{\partial P_{bb}}\frac{\text{d} P_{bb}}{\text{d} t}.
\end{equation}

For simplicity we choose the bias parameters such that $S_{aa} = s_a$ and $S_{ab} = 1-s_a$ and similarly for group $b$. For consistency we always choose to call the smaller group group $a$ and the bigger group group $b$. Since we only have two classes the sizes of the groups are linked $n_a+n_b = 1$.

\subsubsection{Fixed points and stability}
The rate equations~\eqref{eq:rateT} are non-linear equations
which in general do not have closed form solutions, but can still be analysed with standard techniques in dynamical systems theory.
We start by considering the fixed points of the system. The fixed points can be solved by simultaneously solving the equations 
\begin{equation}
\label{eq:fixedpoints}
\begin{cases}
\frac{\text{d} T_{aa}}{\text{d} t} = 0 \\
\frac{\text{d} T_{bb}}{\text{d} t} = 0,
\end{cases}
\end{equation}
where we must exclude $(T_{aa},T_{bb}) = (1, 1)$, due to constraints on~\eqref{eq:rateT}. This means that we are possibly excluding a line of fixed points that have $P_{aa} = 1-P_{bb}$. By plugging this into the fixed point equation for $P$ we get\\
\begin{equation}
\begin{cases}
\left.\frac{\text{d} P_{aa}}{\text{d}t}\right|_{P_{aa} = 1-P_{bb}} = (1-c)(1-n_a)n_a(1-s_a) = 0\\
\left.\frac{\text{d} P_{bb}}{\text{d}t}\right|_{P_{aa} = 1-P_{bb}} = (1-c)n_a(1-n_a)(1-s_b) = 0,
\end{cases}
\end{equation}
which has the solutions $c = 1$, $n_a = 1$, $n_a=0$, and $s_a = 1$ and $s_b=1$. This means that $P_{aa} = 1-P_{bb}$ is a line of fixed points only in the extreme cases when there is only triadic closure, only one group or when both groups are only willing to connect to the same group.

By setting values for the four parameters of the system ($n_a$, $c$, $s_a$ and $s_b$) we could solve the fixed points numerically. However, when we leave one of the parameters free we can solve the equations symbolically (which we have done using symbolic computations with Mathematica~\cite{SI_Mathematica}). The solutions will be roots of single variable polynomial equations. %

We use linear stability analysis to find the local stability of the fixed points. Let
\begin{equation}
A = \left.\begin{pmatrix}
\frac{\partial f}{\partial T_{aa}} & \frac{\partial f}{\partial T_{bb}}\\
\frac{\partial g}{\partial T_{bb}} & \frac{\partial g}{\partial T_{bb}}
\end{pmatrix}\right|_{(T^*_{aa},T^*_{bb})}
\end{equation}
 be the Jacobian matrix of the system 
\begin{equation}
\begin{cases}
\frac{\delta T_{aa}}{\delta t} = f(T_{aa},T_{bb})\\
\frac{\delta T_{bb}}{\delta t} = g(T_{aa},T_{bb})
\end{cases}
\end{equation} 
at a fixed point $(T^*_{aa},T^*_{bb})$ and where $f$ and $g$ are the functions determined by Eq. \ref{eq:rateT}. By calculating the trace and determinant of the Jacobian matrix we can then determine the type (node, spiral, etc.) and stability (stable, unstable, saddle) of the fixed points.

By solving the system for several sets of the parameters (see Fig. 2E in the main text) we find that there are always either one, three or five fixed points. If there is only one fixed point, it is a stable node. If there are three fixed points, two of them are stable nodes and the one between them (in the $(T_{aa},T_{bb})$-plane) is a saddle node. When there are five fixed points, three of them are stable nodes and the two in between the stable nodes are saddle nodes. The only exceptions to this are the extreme cases where $P_{aa} = 1-P_{bb}$ is a fixed point as discussed above.

\subsubsection{Edge cases}
\begin{figure}[h]
\centering
\includegraphics[scale = 1]{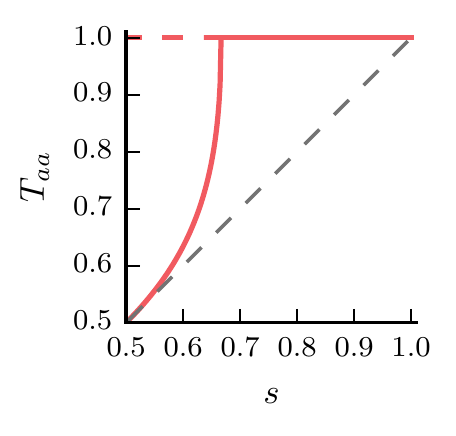}
\caption{The red line shows the stable (solid line) and unstable (dashed line) fixed points of the system when $c=1$.
 The dashed black line corresponds to the $c=0$ system.}
\label{fig:c1fig}
\end{figure}
If triadic closure is the only mechanism, i.e. $c=1$, and the groups have the same homophily biases, $s_a = s_b = s$, the rate equation~\eqref{eq:rateP} simplifies to
\begin{equation}
\begin{split}
n_a(T^2)_{aa} -n_aT_{aa}\left(s(T^2)_{aa} + (1-s)(T^2)_{ab}\right)=0\\
n_b(T^2)_{bb} -n_bT_{bb}\left(s(T^2)_{bb} + (1-s)(T^2)_{ba}\right)=0
\end{split}
\end{equation}
which can be solved in closed form. %
In terms of the $P$-matrix we obtain
\begin{equation}
P_{aa} = \begin{cases}
1 - P_{bb} &\\
\frac{s - \sqrt{s(2-3s)}}{4(-1+2s)}, & \mbox{if } 0 \leq s < 2/3\,,
\end{cases}
\end{equation}
where in the latter case $P_{bb}=P_{aa}$. We exclude the solution $P_{aa} = P_{bb} = \frac{s + \sqrt{s(2-3s)}}{4(-1+2s)}$, because it would lead to $P_{aa} + P_{bb} > 1$.

Figure~\ref{fig:c1fig} shows the fixed points of the system in terms of $T_{aa}$. The solution $T_{aa} = T_{bb} = 1$, corresponding to $P_{aa} = 1-P_{bb}$, is an unstable center ($\text{Det} (J) = 0$, $\text{Tr} (J) > 0$) when $s < 2/3$ and becomes stable at $s=2/3$. The transition point $s = 2/3$ is not smooth but it is continuous. Note that the behaviour of the system with only triadic closure is independent of the group sizes.

In the other extreme case, when one group only accepts links to the same group, i.e. $s_{a}=1$ and $s_{b} \neq 1$ (or $s_b=1$ and $s_a \neq 1$), numerical treatment (where $s_{a} = 1$  (or $s_b=1$) and $s_b (\text{or } s_a), n_a, c \in \{0.02, 0.04, \dots, 1\}$) shows that the system does not have any fixed points unless $c=1$, $n_{a} = 1$ (or $n_b=1$) or $s_a = s_b =1$. 

\subsubsection{Classification of the fixed points}
\begin{figure}[h]
\centering
\includegraphics[width = 0.44 \textwidth]{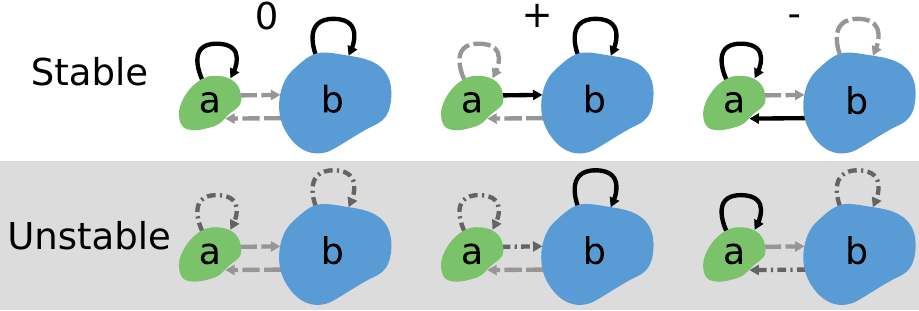}\\
\includegraphics[width = 0.44 \textwidth]{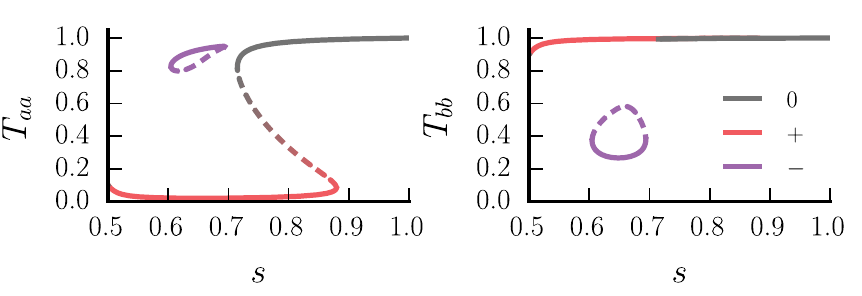}
\caption{(Top) Schematic illustrations of the different classes of fixed points and (bottom) bifurcation diagrams for a system with model parameters $n_a = 0.1$, $c = 0.95$ and $s_a = s_b = s$. Different colors correspond to different bifurcation branches. Solid lines correspond to stable fixed points and dotted lines to saddle points. Note that the three branches that are seen in the $T_{aa}$ plot around $s= 0.7$ and $s=0.9$ are almost overlapping in the $T_{bb}$-plot.}
\end{figure}

The only type of bifurcation we see (as we vary any of the four parameters) in the system is an imperfect pitchfork bifurcation, meaning that at the bifurcation point one fixed point splits into three different branches. The bifurcation is imperfect since the two new fixed points may be far from the original fixed point. Our classification of the fixed points is based on these bifurcations. 

Since the fixed points lie in the $(T_{aa},T_{bb})$-plane, we have three possible types of branches that emerge in the bifurcation point, either the branch is in the upper left half of the $(T_{aa},T_{bb})$-plane (corresponding to the + configuration, where the larger group forms the core), the branch is in the lower left half (corresponding to the - configuration, where the smaller group forms the core) or the branch is closer to the $T_{aa} = T_{bb}$ -line (corresponding to the 0 configurations, where both groups are homophilous and their homophily can be amplified in both groups). 

When there are three or five fixed points we can classify the stable points by considering the positions of the fixed points in the $(T_{aa},T_{bb})$-plane. The branches with the saddle points are harder to classify, as they may connect two different stable branches and thus change configuration as we vary the parameters (see Fig 2.). However, the ends of the saddle branches can be classified into the configurations corresponding to the stable branches that they connect to. When there is only one fixed point we determine the configuration by considering which branch the point belongs to in the bifurcation diagram.

Note that our classification is purely based on bifurcations and the different branches that emerge. Along a single branch one group may change from being homophilous to heterophilous or vice versa, however, the core-periphery structure remains in any given branch. Other ways to classify the fixed points could involve comparing the homophilies of the groups using an index (below we introduce a useful index based on our model), or whether or not triadic closure amplifies the homophily of the groups.

\subsubsection{Phase diagrams}
Our dynamical system has 4 parameters and a full summarization of this 4 dimensional space would be cumbersome. Instead, we focus on 2 dimensional slices chosen from architypical places in the space (Fig 2E in the main text). 
To find the 2 dimensional phase diagrams of the system we first set two of the parameters to some values, e.g. $n_a = 0.1$ and $c=0.9$. We then solve the system in a grid of the two other parameters, e.g. a 100x100 grid for $s_a$ and $s_b$. We look at the number of fixed points in each of the grid points. If the number of fixed points changes between two points we know that a boundary of a region (bifurcation point) must exist between those points.

After we have found the rough boundaries in the grid we use the bisection method to find the boundaries to a desired accuracy (we set the tolerance to $10^{-4}$) first in one direction of the grid and then in the other.

\subsection{Simulations}
\subsubsection{Accuracy of the mean field approach}
We use simulations to verify the predictions of the mean field equations. To see how well the predicted stable fixed points correspond to the fixed points in the simulation we initially set the network to be in a stable fixed point in terms of the transition matrix $T$ (but otherwise maximally random). We then run the simulation for 1000 steps and calculate the distance to the fixed points predicted by the mean field. Simulations were run for networks with $N=10\,000$ %
nodes and average degree $\langle k \rangle = 50$.

\begin{figure*}[t]
\centering
\includegraphics[width = 0.49\textwidth]{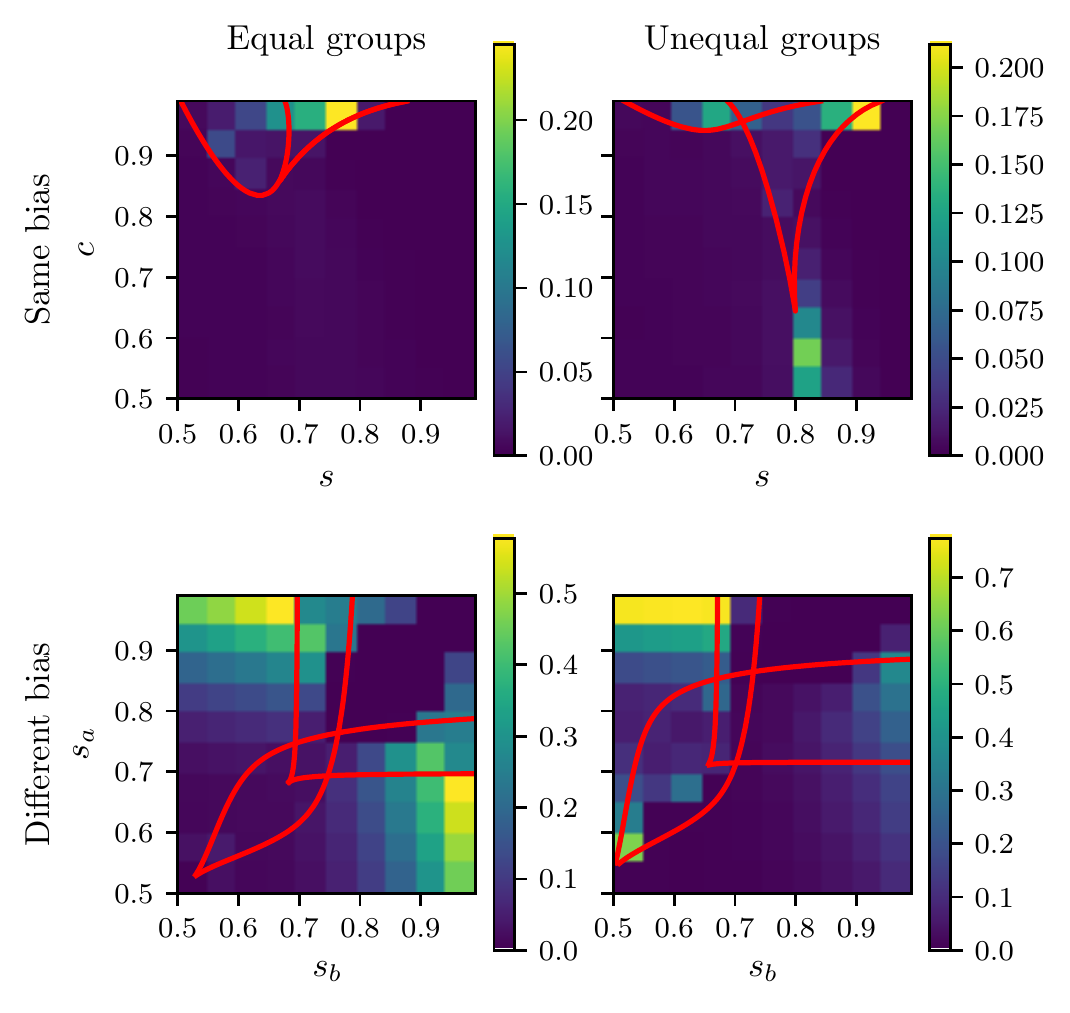} 
\includegraphics[width = 0.49\textwidth]{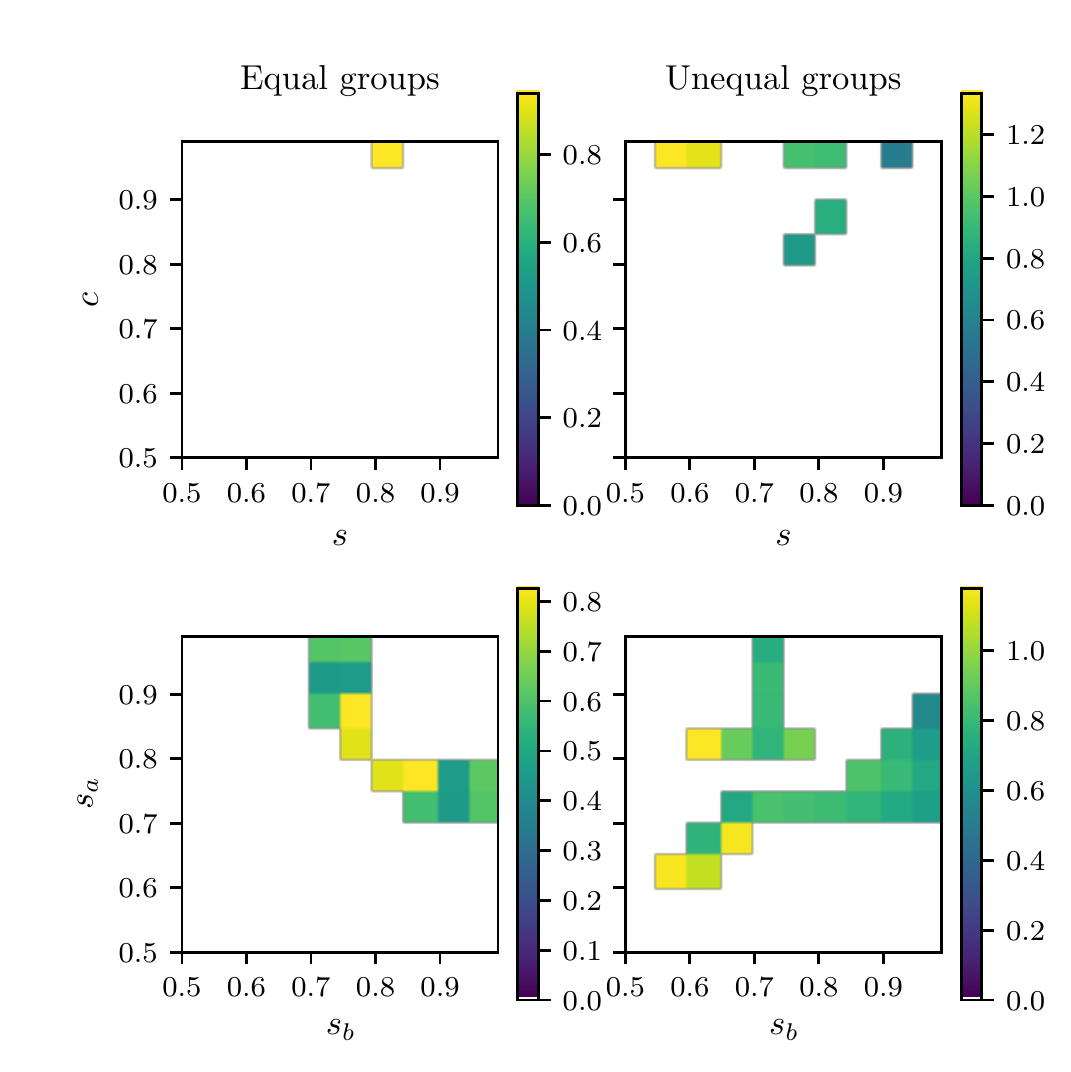}
\caption{(Left) Distance to the nearest fixed point predicted by the mean field equation after 1000 simulation steps averaged over 100 simulations. (Right) Distance between the initial state (a predicted fixed point) and the final state of the system for those runs where the nearest fixed point after the simulation isn't the same as the initial state. In both panels the parameters correspond to those in Fig. 2E in the main text, i.e. in the upper row $s_a= s_b=s$ and in the lower row $s_a \neq s_b$ and $c=0.95$, for equal group sizes $n_a=n_b$ (left) and unequal group sizes $n_a = 0.1$ (right).
Note that the maximum possible distance between two points in the $(T_{aa}, T_{bb})$-plane is $\sqrt{2}\approx 1.41$.}
\label{fig:accuracy}
\end{figure*}

The left panel in figure~\ref{fig:accuracy} shows a heat map of the distance to the nearest fixed point predicted by the mean field equation for several sets of parameters. The mean field prediction is most accurate when $s_a \approx s_b$ and when there is only one fixed point in the system. For some of the parameter sets, however, the nearest fixed point after the simulation was different from the initial state. The right panel shows the distance between those points. In these cases the system ends up far away from the initial state. 
In some cases the final state of the systems is very close to a fixed point which is not the initial one. 
This suggests that the system goes to a fixed point predicted by the mean field method, but just not the one that we anticipated. This could be because \emph{local} stability analysis is sensitive to the random fluctuations of the simulations which might carry the system state to a different fixed point.  %

\subsubsection{Time evolution and trajectories}
\begin{figure}
\includegraphics[scale=1]{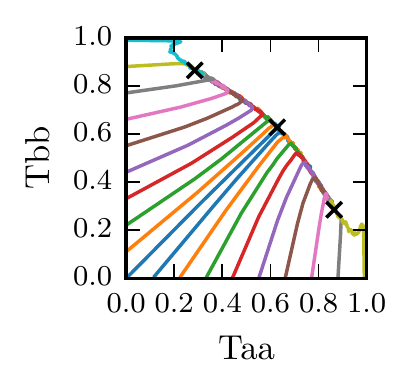}
\includegraphics[scale=1]{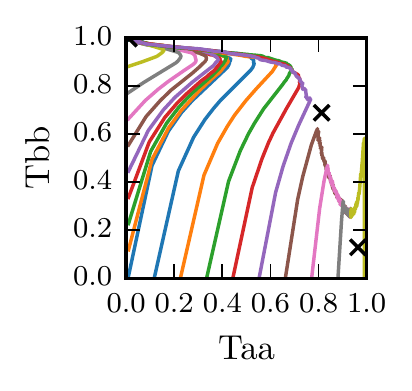}
\caption{Simulated trajectories for a system with equal group sizes (left, $n_a = 0.5$, $s_a=s_b = 0.6$, $c=0.9$) and unequal group sizes (right, $n_a = 0.1$, $s_a=s_b=0.65$, $c = 0.97$). Black crosses show the analytical fixed points obtained from the mean field equation.}
\label{fig:tra}
\end{figure}

To investigate the global dynamics of the triadic closure system we have simulated the time evolution of the system. In these simulations the networks have $N=10\,000$ nodes and average degree $\langle k \rangle = 20$. We find that the trajectories first go rapidly to an area between the nullclines and then evolve slowly towards the fixed points. This slowing down of the system is not surprising since near the nullclines $\dot{T}_{aa} \approx 0$ and $\dot{T}_{bb} \approx 0$, since the right hand side of the Master equation is continuous. Figure~\ref{fig:tra} shows some trajectories for two systems with either equal or unequal group sizes, high triadic closure probability and moderate homophily bias.

\begin{figure}
\includegraphics[scale=1]{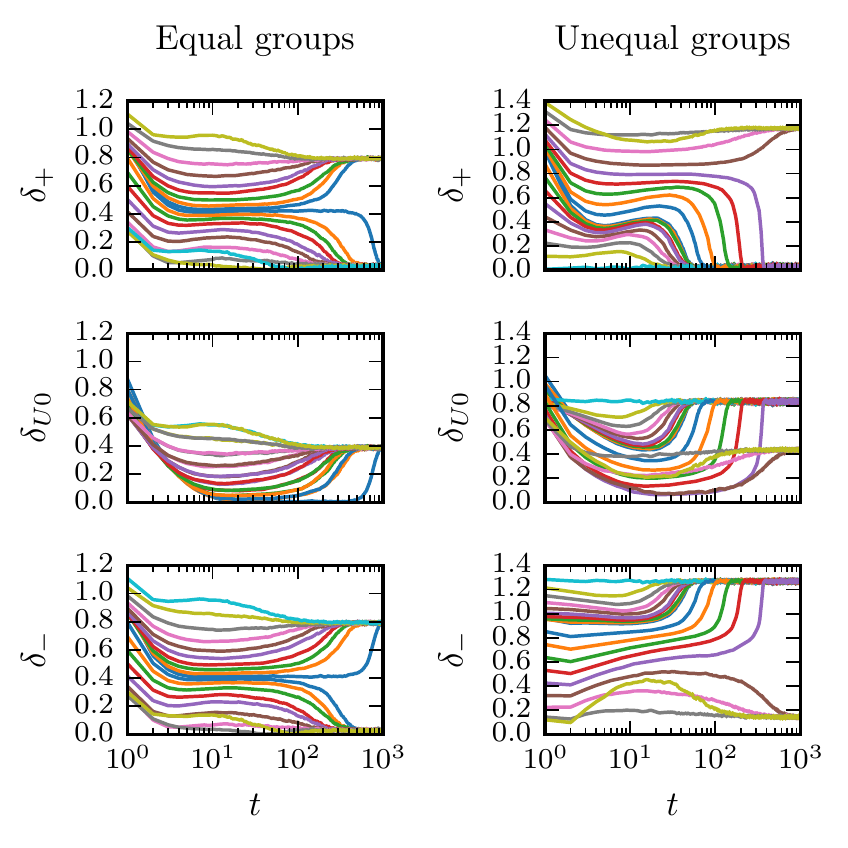}
\caption{Time evolution of two systems as measured by the Euclidean distance to each of the fixed points for different initial conditions. The system is initially in a state where either $T_{aa}=0.01$ or $T_{bb}=0.01$ (see Fig. \ref{fig:tra}). The system has two very different time scales. First it evolves towards the nullclines rapidly ($\sim$ few Monte Carlo steps). After entering the area between the nullclines the system starts to slowly ($\sim$ few hundred Monte Carlo steps) evolve towards the stable fixed points. The model parameters are $n_a = 0.5, s_a = s_b = 0.6$ and $c = 0.9$ in the left panel and $n_a = 0.1$, $s_a=s_b = 0.65$ and $c=0.97$ in the right panel.}
\label{fig:globdyn}
\end{figure}

Figure~\ref{fig:globdyn} shows the time evolution of the trajectories measured by the Euclidean distance to each of the fixed points for two systems. In the early stages of the simulation the distance to all of the fixed points decreases rapidly. After the fast initial stage the trajectories slowly start going towards the stable fixed points. The separate blue line in the left panel of figure~\ref{fig:globdyn} corresponds to a trajectory that goes through the saddle point, this system spends a long time near that point until random fluctuations drive it to one of the stable points.

\subsection{Data analysis}
\subsubsection{Details of the datasets}
Here we give further details of the 6 network data sets we use in the main text.
\begin{itemize}
\item \emph{Mobile phone network}~\cite{SI_Onnela2007Mobile} is based on call data records from 2009 and customer information database of a single mobile service provider in a European country. The network is constructed based on this data in a way that each subscriber becomes a node.
 Two individuals have a link between them if both of them called the other at least once during the period of 12 months. We split the network into two groups according to the reported gender of the individuals. We exclude those individuals that have no gender information in the database.

\item A network of \emph{political blogs}~\cite{SI_Adamic2005polblogs} from February 2005, where blogs are classified as either liberal or conservative. A link exists between two blogs if either one has a link to the other on the front page.

\item \emph{last.fm} is a web service where users can share their musical tastes and assign other users as friends. We use the mutualized network of friendships and classify each user by gender. In the website the users can include their gender in their profile information, but when the self-reported gender is not available it is inferred from the real first name reported by the user.

\item The \emph{facebook100}~\cite{SI_Traud2012facebook100} dataset contains the Facebook networks of 100 universities in 2005. We can estimate the homophily bias in each of the 100 universities in the facebook100 dataset. We took the two largest classes for each university and took the subnetwork of containing those classes. See table~\ref{tab:data_large} for properties of these networks.

\item The \emph{primary school}~\cite{SI_Stehle2011Sociopatterns} dataset contains the aggregated contact network of primary school students and their teachers during two consecutive days. The contacts were recorded in 20s intervals. To better reflect the more ``meaningful'' contacts we take only those contacts that had a combined duration of over 20 minutes during the two days. The students are split into groups according to their gender. As teachers did not have gender information they are excluded from the network.
\end{itemize}

\begin{table*}[t]
\centering
\caption{\textbf{Properties of the Facebook datasets.} \textnormal{List of the Facebook social networks used in this study, their main properties and estimated model parameters. $N$ is the number of nodes in the network, $\langle k \rangle$ the average degree, $n_a$ the fraction of nodes in the smaller attribute group, $\text{HI}_{a} = (T_{aa}-n_a)/(1-n_a)$ and $\text{HI}_{b} = (T_{bb}-n_b)/(1-n_b)$ the Coleman homophily indices of the groups, $s_a(0)$ and $s_b(0)$ the estimates of the bias parameters when $c=0$, and $s_a(1)$ and $s_b(1)$ the estimates when $c=1$.}}
\label{tab:data_large}
\begin{tabular}{l cccccc|cccc}
 & class $a$, class $b$ & $N$ & $\langle k \rangle$ & $n_a$ & $\text{HI}_a$ & $\text{HI}_b$ &$ s_a(0) / s_b(0)$ & $s_a(1) / s_b (1)$ \\
\hline
American75 & 2008, 2007 & 2313 & 60.07 & 48.9\% & 0.412 & 0.467 & 0.709/0.731 & 0.639/0.630 \\
Amherst41 & 2008, 2009 & 756 & 56.13 & 49.7\% & 0.847 & 0.917 & 0.924/0.958 & 0.670/0.662 \\
Auburn71 & 2009, 2008 & 6022 & 60.71 & 46.9\% & 0.640 & 0.541 & 0.827/0.763 & 0.649/0.656 \\
BC17 & 2008, 2009 & 4250 & 65.98 & 49.0\% & 0.906 & 0.927 & 0.954/0.963 & 0.668/0.665 \\
BU10 & 2008, 2009 & 7371 & 48.72 & 47.0\% & 0.872 & 0.908 & 0.939/0.952 & 0.669/0.663 \\
Baylor93 & 2009, 2008 & 4726 & 69.07 & 49.9\% & 0.842 & 0.794 & 0.921/0.897 & 0.662/0.667 \\
Berkeley13 & 2008, 2007 & 7546 & 63.79 & 49.6\% & 0.539 & 0.579 & 0.770/0.789 & 0.652/0.646 \\
Bingham82 & 2007, 2008 & 3669 & 53.30 & 48.6\% & 0.432 & 0.415 & 0.719/0.704 & 0.633/0.631 \\
Bowdoin47 & 2008, 2009 & 826 & 54.87 & 49.9\% & 0.873 & 0.931 & 0.937/0.965 & 0.669/0.663 \\
Brandeis99 & 2008, 2007 & 1410 & 65.81 & 48.6\% & 0.657 & 0.688 & 0.832/0.841 & 0.661/0.655 \\
Brown11 & 2008, 2006 & 2692 & 70.14 & 48.5\% & 0.774 & 0.803 & 0.889/0.899 & 0.666/0.661 \\
Bucknell39 & 2008, 2009 & 1609 & 66.12 & 49.8\% & 0.818 & 0.866 & 0.909/0.933 & 0.668/0.662 \\
Cal65 & 2008, 2007 & 3339 & 45.59 & 48.3\% & 0.499 & 0.646 & 0.754/0.819 & 0.662/0.640 \\
Caltech36 & 2008, 2006 & 322 & 35.90 & 46.6\% & 0.413 & 0.602 & 0.715/0.793 & 0.661/0.629 \\
Carnegie49 & 2009, 2008 & 2222 & 46.02 & 49.3\% & 0.824 & 0.766 & 0.913/0.882 & 0.661/0.667 \\
Colgate88 & 2009, 2008 & 1278 & 61.92 & 49.9\% & 0.883 & 0.916 & 0.942/0.958 & 0.668/0.664 \\
Columbia2 & 2006, 2008 & 3593 & 61.69 & 49.5\% & 0.762 & 0.804 & 0.882/0.901 & 0.666/0.661 \\
Cornell5 & 2007, 2008 & 5794 & 77.55 & 49.5\% & 0.673 & 0.626 & 0.838/0.812 & 0.654/0.659 \\
Dartmouth6 & 2008, 2007 & 2098 & 79.62 & 49.0\% & 0.708 & 0.713 & 0.856/0.855 & 0.661/0.659 \\
Duke14 & 2008, 2009 & 3232 & 76.06 & 49.9\% & 0.828 & 0.891 & 0.914/0.946 & 0.669/0.662 \\
Emory27 & 2009, 2008 & 2729 & 66.23 & 49.0\% & 0.855 & 0.816 & 0.929/0.907 & 0.663/0.667 \\
FSU53 & 2008, 2009 & 8870 & 43.78 & 48.9\% & 0.711 & 0.769 & 0.858/0.883 & 0.666/0.658 \\
GWU54 & 2008, 2009 & 4420 & 59.04 & 47.7\% & 0.885 & 0.913 & 0.945/0.955 & 0.668/0.664 \\
Georgetown15 & 2008, 2007 & 2929 & 86.13 & 47.6\% & 0.686 & 0.708 & 0.848/0.849 & 0.663/0.656 \\
Hamilton46 & 2008, 2009 & 868 & 57.65 & 49.2\% & 0.881 & 0.915 & 0.941/0.957 & 0.668/0.664 \\
Harvard1 & 2007, 2008 & 3432 & 109.92 & 50.0\% & 0.735 & 0.732 & 0.868/0.866 & 0.661/0.662 \\
Haverford76 & 2008, 2009 & 584 & 61.07 & 49.3\% & 0.695 & 0.791 & 0.849/0.894 & 0.668/0.656 \\
Howard90 & 2008, 2007 & 1903 & 91.24 & 48.9\% & 0.616 & 0.631 & 0.811/0.813 & 0.657/0.653 \\
Indiana69 & 2008, 2009 & 10028 & 63.40 & 49.3\% & 0.767 & 0.783 & 0.885/0.890 & 0.664/0.662 \\
JMU79 & 2009, 2008 & 5856 & 55.36 & 46.0\% & 0.838 & 0.819 & 0.924/0.903 & 0.665/0.664 \\
Johns Hopkins55 & 2008, 2009 & 1823 & 44.88 & 49.7\% & 0.812 & 0.897 & 0.907/0.948 & 0.671/0.660 \\
Lehigh96 & 2009, 2008 & 1983 & 63.19 & 46.0\% & 0.894 & 0.880 & 0.951/0.936 & 0.666/0.666 \\
MIT8 & 2008, 2007 & 1870 & 70.51 & 46.6\% & 0.490 & 0.542 & 0.754/0.763 & 0.652/0.638 \\
MSU24 & 2008, 2009 & 10828 & 44.09 & 46.9\% & 0.670 & 0.770 & 0.842/0.880 & 0.670/0.653 \\
MU78 & 2008, 2009 & 6012 & 60.82 & 48.1\% & 0.861 & 0.887 & 0.933/0.941 & 0.668/0.664 \\
Maine59 & 2008, 2007 & 3511 & 37.95 & 46.4\% & 0.294 & 0.377 & 0.655/0.680 & 0.626/0.606 \\
Maryland58 & 2008, 2007 & 7689 & 63.10 & 48.5\% & 0.508 & 0.568 & 0.757/0.780 & 0.653/0.642 \\
Mich67 & 2008, 2007 & 1245 & 27.45 & 46.6\% & 0.349 & 0.414 & 0.683/0.699 & 0.634/0.617 \\
Michigan23 & 2008, 2009 & 10390 & 51.31 & 48.4\% & 0.773 & 0.848 & 0.889/0.922 & 0.670/0.659 \\
Middlebury45 & 2008, 2007 & 1104 & 74.69 & 48.4\% & 0.679 & 0.681 & 0.843/0.837 & 0.660/0.657 \\
Mississippi66 & 2007, 2008 & 3397 & 85.63 & 49.8\% & 0.429 & 0.358 & 0.715/0.678 & 0.623/0.631 \\
NYU9 & 2008, 2009 & 7890 & 53.43 & 48.9\% & 0.866 & 0.896 & 0.934/0.947 & 0.668/0.664 \\
Northeastern19 & 2008, 2009 & 4671 & 44.18 & 49.3\% & 0.654 & 0.627 & 0.829/0.812 & 0.655/0.657 \\
Northwestern25 & 2008, 2009 & 3556 & 61.98 & 48.8\% & 0.806 & 0.882 & 0.905/0.940 & 0.670/0.660 \\
Notre Dame57 & 2008, 2007 & 4345 & 86.46 & 48.4\% & 0.640 & 0.645 & 0.824/0.819 & 0.658/0.655 \\
Oberlin44 & 2008, 2007 & 1097 & 51.59 & 46.6\% & 0.442 & 0.509 & 0.730/0.746 & 0.648/0.631 \\
Oklahoma97 & 2008, 2007 & 5685 & 75.16 & 47.5\% & 0.272 & 0.443 & 0.641/0.716 & 0.634/0.608 
\end{tabular}
\end{table*}

\begin{table*}[t]
\centering
\begin{tabular}{l cccccc|cccc}
 & class $a$, class $b$ & $N$ & $\langle k \rangle$ & $n_a$ & $\text{HI}_a$ & $\text{HI}_b$ &$ s_a(0) / s_b(0)$ & $s_a(1) / s_b (1)$ \\
\hline
Penn94 & 2008, 2009 & 14251 & 42.49 & 47.2\% & 0.768 & 0.794 & 0.889/0.893 & 0.666/0.660 \\
Pepperdine86 & 2008, 2006 & 1356 & 68.18 & 47.2\% & 0.505 & 0.594 & 0.760/0.790 & 0.657/0.640 \\
Princeton12 & 2008, 2009 & 2112 & 56.49 & 49.5\% & 0.845 & 0.911 & 0.923/0.955 & 0.670/0.662 \\
Reed98 & 2008, 2007 & 372 & 35.45 & 44.4\% & 0.380 & 0.435 & 0.704/0.704 & 0.641/0.619 \\
Rice31 & 2008, 2007 & 1372 & 73.93 & 48.4\% & 0.435 & 0.454 & 0.722/0.723 & 0.639/0.632 \\
Rochester38 & 2008, 2009 & 1761 & 47.96 & 48.7\% & 0.824 & 0.854 & 0.914/0.925 & 0.667/0.663 \\
Rutgers89 & 2008, 2007 & 8875 & 51.61 & 49.2\% & 0.519 & 0.602 & 0.762/0.799 & 0.656/0.644 \\
Santa74 & 2008, 2006 & 1512 & 67.45 & 44.8\% & 0.671 & 0.726 & 0.847/0.853 & 0.667/0.653 \\
Simmons81 & 2008, 2007 & 636 & 37.27 & 44.7\% & 0.592 & 0.647 & 0.810/0.811 & 0.663/0.646 \\
Smith60 & 2008, 2006 & 1158 & 41.70 & 46.0\% & 0.496 & 0.659 & 0.758/0.821 & 0.666/0.637 \\
Stanford3 & 2008, 2006 & 3191 & 90.07 & 49.0\% & 0.724 & 0.744 & 0.864/0.870 & 0.663/0.660 \\
Swarthmore42 & 2008, 2009 & 600 & 45.67 & 48.0\% & 0.754 & 0.893 & 0.881/0.944 & 0.674/0.656 \\
Syracuse56 & 2009, 2008 & 5075 & 54.76 & 46.8\% & 0.822 & 0.786 & 0.916/0.888 & 0.663/0.665 \\
Temple83 & 2008, 2007 & 4967 & 44.60 & 45.2\% & 0.523 & 0.631 & 0.774/0.805 & 0.663/0.640 \\
Tennessee95 & 2008, 2009 & 5992 & 52.71 & 45.2\% & 0.589 & 0.696 & 0.807/0.838 & 0.668/0.646 \\
Texas80 & 2008, 2006 & 10241 & 49.47 & 47.6\% & 0.571 & 0.674 & 0.791/0.832 & 0.664/0.646 \\
Texas84 & 2008, 2007 & 12889 & 64.14 & 49.2\% & 0.389 & 0.441 & 0.696/0.719 & 0.635/0.627 \\
Trinity100 & 2008, 2009 & 974 & 59.85 & 48.9\% & 0.886 & 0.927 & 0.944/0.962 & 0.669/0.664 \\
Tufts18 & 2008, 2007 & 2257 & 77.10 & 49.1\% & 0.626 & 0.639 & 0.815/0.818 & 0.657/0.654 \\
Tulane29 & 2008, 2007 & 2713 & 66.75 & 48.7\% & 0.523 & 0.567 & 0.765/0.780 & 0.652/0.644 \\
UC33 & 2008, 2007 & 6172 & 55.91 & 48.6\% & 0.651 & 0.706 & 0.829/0.850 & 0.663/0.654 \\
UC61 & 2008, 2007 & 4826 & 56.99 & 45.1\% & 0.591 & 0.690 & 0.808/0.835 & 0.667/0.646 \\
UC64 & 2008, 2007 & 3156 & 40.75 & 42.2\% & 0.515 & 0.601 & 0.778/0.783 & 0.664/0.636 \\
UCF52 & 2008, 2009 & 5869 & 34.42 & 49.5\% & 0.690 & 0.725 & 0.846/0.861 & 0.663/0.658 \\
UCLA26 & 2007, 2006 & 6748 & 56.58 & 49.9\% & 0.321 & 0.460 & 0.661/0.730 & 0.635/0.619 \\
UCSB37 & 2008, 2007 & 5477 & 60.42 & 46.9\% & 0.643 & 0.682 & 0.829/0.834 & 0.663/0.653 \\
UCSC68 & 2008, 2007 & 3707 & 47.94 & 46.2\% & 0.579 & 0.656 & 0.799/0.820 & 0.663/0.646 \\
UCSD34 & 2008, 2007 & 5584 & 57.40 & 45.0\% & 0.636 & 0.672 & 0.830/0.825 & 0.664/0.651 \\
UChicago30 & 2008, 2009 & 2034 & 37.89 & 47.6\% & 0.809 & 0.874 & 0.908/0.934 & 0.670/0.660 \\
UConn91 & 2008, 2009 & 6214 & 44.24 & 48.0\% & 0.779 & 0.789 & 0.893/0.891 & 0.665/0.662 \\
UF21 & 2009, 2007 & 11337 & 44.39 & 47.5\% & 0.775 & 0.800 & 0.892/0.896 & 0.666/0.661 \\
UGA50 & 2009, 2007 & 7817 & 53.81 & 49.5\% & 0.803 & 0.824 & 0.902/0.912 & 0.666/0.663 \\
UIllinois20 & 2008, 2009 & 11602 & 56.39 & 48.4\% & 0.757 & 0.839 & 0.881/0.917 & 0.670/0.658 \\
UMass92 & 2008, 2007 & 6475 & 51.51 & 46.3\% & 0.444 & 0.518 & 0.732/0.750 & 0.650/0.631 \\
UNC28 & 2008, 2009 & 6656 & 55.80 & 49.8\% & 0.672 & 0.785 & 0.836/0.892 & 0.669/0.654 \\
UPenn7 & 2008, 2009 & 4528 & 63.38 & 49.1\% & 0.871 & 0.923 & 0.937/0.961 & 0.669/0.663 \\
USC35 & 2007, 2006 & 6041 & 71.29 & 48.0\% & 0.299 & 0.412 & 0.654/0.701 & 0.629/0.611 \\
USF51 & 2008, 2007 & 4518 & 32.69 & 43.7\% & 0.266 & 0.442 & 0.646/0.706 & 0.639/0.602 \\
USFCA72 & 2008, 2007 & 1094 & 47.92 & 46.8\% & 0.581 & 0.617 & 0.798/0.801 & 0.659/0.648 \\
UVA16 & 2008, 2007 & 5889 & 83.05 & 49.2\% & 0.585 & 0.625 & 0.794/0.811 & 0.657/0.650 \\
Vanderbilt48 & 2008, 2009 & 2923 & 75.78 & 49.1\% & 0.873 & 0.906 & 0.938/0.952 & 0.668/0.664 \\
Vassar85 & 2008, 2009 & 1171 & 51.44 & 49.6\% & 0.809 & 0.868 & 0.905/0.934 & 0.669/0.661 \\
Vermont70 & 2009, 2008 & 3173 & 39.93 & 48.6\% & 0.872 & 0.824 & 0.937/0.910 & 0.663/0.668 \\
Villanova62 & 2008, 2009 & 2877 & 70.42 & 47.2\% & 0.885 & 0.917 & 0.945/0.956 & 0.669/0.664 \\
Virginia63 & 2009, 2008 & 7883 & 41.58 & 49.8\% & 0.759 & 0.707 & 0.880/0.853 & 0.658/0.665 \\
Wake73 & 2008, 2009 & 2068 & 68.63 & 48.3\% & 0.824 & 0.873 & 0.914/0.935 & 0.669/0.662 \\
WashU32 & 2008, 2007 & 2721 & 90.04 & 48.0\% & 0.648 & 0.684 & 0.829/0.838 & 0.662/0.654 \\
Wellesley22 & 2008, 2006 & 1041 & 49.22 & 47.0\% & 0.732 & 0.782 & 0.872/0.886 & 0.667/0.658 \\
Wesleyan43 & 2008, 2009 & 1293 & 52.11 & 47.8\% & 0.833 & 0.900 & 0.919/0.948 & 0.670/0.661 \\
William77 & 2008, 2009 & 2418 & 58.34 & 48.6\% & 0.830 & 0.875 & 0.917/0.936 & 0.668/0.662 \\
Williams40 & 2008, 2009 & 941 & 52.86 & 48.0\% & 0.842 & 0.898 & 0.924/0.947 & 0.670/0.662 \\
Wisconsin87 & 2008, 2007 & 8072 & 55.78 & 46.7\% & 0.460 & 0.539 & 0.739/0.762 & 0.652/0.634 \\
Yale4 & 2008, 2007 & 2596 & 94.48 & 49.4\% & 0.650 & 0.698 & 0.827/0.848 & 0.662/0.655
\end{tabular}
\end{table*}

\subsubsection{Estimation of intrinsic bias}
To find out the possible model parameters for the empirical networks we can solve the fixed point equations for the bias parameters $s_a$ and $s_b$. The fixed point equations are linear in $s_a$ and $s_b$ and thus they can be solved in closed form such that,
\begin{widetext}
\begin{equation}
\begin{split}
s_a &= \frac{n_bT_{aa} - c T_{aa} [n_b - (1 - T_{aa})(T_{aa} + T_{bb})]}{n_a(1 - T_{aa}) + n_b T_{aa} -
  c (-1 + 2 T_{aa}) [n_b - (1 - T_{aa})(T_{aa} + T_{bb})]} \quad \text{and}\\
s_b &= \frac{n_aT_{bb}-cT_{bb}[n_a-(1-T_{bb})(T_{aa}+T_{bb})]}{n_b(1-T_{bb})+n_aT_{bb}-c(-1+2T_{bb})[n_a-(1-T_{bb})(T_{aa}+T_{bb})]},
\end{split}
\label{eq:sasb}
\end{equation}
\end{widetext}
where $n_b=1-n_a$.

We can measure the sizes of the groups $n_a$ and $n_b$ as well as the proportion of intra-group links $T_{aa}$ and $T_{bb}$ directly from the data. We can then calculate the possible values for $s_a$ and $s_b$ for a given triadic closure probability $c$. Specially we can obtain a range of possible bias values since $c$ can only vary between 0 and 1.

When we set $c=0$ we can use equations~\ref{eq:sasb} to obtain a naive estimate of the biases without triadic closure
\begin{equation}
\begin{split}
s_a(c=0) = \frac{n_b T_{aa}}{n_a(1-T_{aa})+ n_b T_{aa}}\\
s_b(c=0) = \frac{n_a T_{bb}}{n_b(1-T_{bb})+ n_a T_{bb}}.
\end{split}
\end{equation}
The correction for the groups sizes done by these estimators are not equivalent to the Coleman's homophily index, because the $s_a$ and $s_b$ values written in terms of the Coleman's index depend also on the group sizes. 


\begin{thebibliography}{10}

\bibitem{Mcpherson2001Birds}
McPherson M, Smith-Lovin L, Cook JM (2001) Birds of a feather: Homophily in
  social networks.
\newblock {\em Annu. Revi. Sociol.} 27(1):415--444.

\bibitem{Aukett1988Gender}
Aukett R, Ritchie J, Mill K (1988) Gender differences in friendship patterns.
\newblock {\em Sex Roles} 19(1-2):57--66.

\bibitem{Shrum1988Friendship}
Shrum W, Cheek~Jr NH, MacD S (1988) Friendship in school: Gender and racial
  homophily.
\newblock {\em Sociol. Educ.} pp. 227--239.

\bibitem{Elkins1993Gender}
Elkins LE, Peterson C (1993) Gender differences in best friendships.
\newblock {\em Sex Roles} 29(7-8):497--508.

\bibitem{Reeder2003Effect}
Reeder HM (2003) The effect of gender role orientation on same-and cross-sex
  friendship formation.
\newblock {\em Sex Roles} 49(3-4):143--152.

\bibitem{Thelwall2008Social}
Thelwall M (2008) Social networks, gender, and friending: An analysis of
  myspace member profiles.
\newblock {\em J. Am. Soc. Inf. Sci. Tec.} 59(8):1321--1330.

\bibitem{Lewis2008Tastes}
Lewis K, Kaufman J, Gonzalez M, Wimmer A, Christakis N (2008) Tastes, ties, and
  time: A new social network dataset using facebook. com.
\newblock {\em Soc. Net.} 30(4):330--342.

\bibitem{Volkovich2014Gender}
Volkovich Y, Laniado D, Kappler KE, Kaltenbrunner A (2014) {\em Gender Patterns
  in a Large Online Social Network}, eds.{} Aiello LM, McFarland D.
\newblock (Social Informatics. SocInfo 2014. Lecture Notes in Computer Science,
  vol 8851, Springer), pp. 139--150.

\bibitem{Zeltzer2015Gender}
Zeltzer D (February 21, 2017) Gender homophily in referral networks:
  Consequences for the medicare physician earnings gap.
\newblock Available at SSRN: \url{https://ssrn.com/abstract=2921482} or
  \url{http://dx.doi.org/10.2139/ssrn.2921482}.

\bibitem{Vicario2016Echo}
Del~Vicario M, et~al. (2016) Echo chambers: Emotional contagion and group
  polarization on facebook.
\newblock {\em Sci. Rep.} 6:37825.

\bibitem{Schmidt2017Anatomy}
Schmidt AL, et~al. (2017) Anatomy of news consumption on facebook.
\newblock {\em P. Natl. Acad. Sci. USA} 114(12):3035--3039.

\bibitem{Dimaggio2012Network}
DiMaggio P, Garip F (2012) Network effects and social inequality.
\newblock {\em Annu. Rev. Sociol.} 38:93--118.

\bibitem{Diprete2006Cumulative}
DiPrete TA, Eirich GM (2006) Cumulative advantage as a mechanism for
  inequality: A review of theoretical and empirical developments.
\newblock {\em Annu. Rev. Sociol.} pp. 271--297.

\bibitem{Currarini2009Economic}
Currarini S, Jackson MO, Pin P (2009) An economic model of friendship:
  Homophily, minorities, and segregation.
\newblock {\em Econometrica} 77(4):1003--1045.

\bibitem{Currarini2016Simple}
Currarini S, Matheson J, Vega-Redondo F (2016) A simple model of homophily in
  social networks.
\newblock {\em Eur. Econ. Rev.} 90:18--39.

\bibitem{Karimi2018Homophily}
Karimi F, G{\'e}nois M, Wagner C, Singer P, Strohmaier M (2018) Homophily
  influences ranking of minorities in social networks.
\newblock {\em Sci. Rep.} 8:11077.

\bibitem{Mcpherson1987Homophily}
McPherson JM, Smith-Lovin L (1987) Homophily in voluntary organizations: Status
  distance and the composition of face-to-face groups.
\newblock {\em Am. Sociol. Rev.} pp. 370--379.

\bibitem{Kossinets2006Empirical}
Kossinets G, Watts DJ (2006) Empirical analysis of an evolving social network.
\newblock {\em Science} 311(5757):88--90.

\bibitem{Kossinets2009Origins}
Kossinets G, Watts DJ (2009) Origins of homophily in an evolving social
  network1.
\newblock {\em Am. J. Sociol.} 115(2):405--450.

\bibitem{Schelling1971Dynamic}
Schelling TC (1971) Dynamic models of segregation.
\newblock {\em J. Math. Sociol.} 1(2):143--186.

\bibitem{Vinkovic2006Physical}
Vinkovi{\'c} D, Kirman A (2006) A physical analogue of the schelling model.
\newblock {\em P. Natl. Acad. Sci. USA} 103(51):19261--19265.

\bibitem{Toivonen2009Comparative}
Toivonen R, et~al. (2009) A comparative study of social network models: Network
  evolution models and nodal attribute models.
\newblock {\em Soc. Net.} 31(4):240--254.

\bibitem{Gong2012Evolution}
Gong NZ, et~al. (2012) Evolution of social-attribute networks: Measurements,
  modeling, and implications using google+ in {\em Proceedings of the 2012
  Internet Measurement Conference}.
\newblock (ACM), pp. 131--144.

\bibitem{Klimek2013Triadic}
Klimek P, Thurner S (2013) Triadic closure dynamics drives scaling laws in
  social multiplex networks.
\newblock {\em New J. Phys.} 15(6):063008.

\bibitem{Bianconi2014Triadic}
Bianconi G, Darst RK, Iacovacci J, Fortunato S (2014) Triadic closure as a
  basic generating mechanism of communities in complex networks.
\newblock {\em Phys. Rev. E} 90(4):042806.

\bibitem{Granovetter1973Strength}
Granovetter MS (1973) The strength of weak ties.
\newblock {\em Am. J. Sociol.} pp. 1360--1380.

\bibitem{Snijders2011Statistical}
Snijders TA (2011) Statistical models for social networks.
\newblock {\em Annu. Rev. Sociol.} 37:131--153.

\bibitem{Kumpula2007Emergence}
Kumpula JM, Onnela JP, Saram{\"a}ki J, Kaski K, Kert{\'e}sz J (2007) Emergence
  of communities in weighted networks.
\newblock {\em Phys. Rev. Lett.} 99(22):228701.

\bibitem{Borgatti2000Models}
Borgatti SP, Everett MG (2000) Models of core/periphery structures.
\newblock {\em Soc. Net.} 21(4):375--395.

\bibitem{Holme2005Core}
Holme P (2005) Core-periphery organization of complex networks.
\newblock {\em Phys. Rev. E} 72(4):046111.

\bibitem{Rombach2017Core}
Rombach P, Porter MA, Fowler JH, Mucha PJ (2017) Core-periphery structure in
  networks (revisited).
\newblock {\em SIAM Rev.} 59(3):619--646.

\bibitem{Traud2012facebook100}
Traud AL, Mucha PJ, Porter MA (2012) Social structure of facebook networks.
\newblock {\em Physica A} 391(16):4165 -- 4180.

\bibitem{Stehle2011Sociopatterns}
Stehlé J, et~al. (2011) High-resolution measurements of face-to-face contact
  patterns in a primary school.
\newblock {\em PLoS ONE} 6(8):e23176.

\bibitem{Adamic2005polblogs}
Adamic LA, Glance N (2005) The political blogosphere and the 2004 u.s.
  election: Divided they blog in {\em Proceedings of the 3rd International
  Workshop on Link Discovery}, LinkKDD '05.
\newblock (ACM, New York, NY, USA), pp. 36--43.

\bibitem{Onnela2007Structure}
Onnela JP, et~al. (2007) Structure and tie strengths in mobile communication
  networks.
\newblock {\em P. Natl. Acad. Sci. USA} 104(18):7332--7336.

\bibitem{Mcpherson1991Evolution}
McPherson JM, Ranger-Moore JR (1991) Evolution on a dancing landscape:
  Organizations and networks in dynamic blau space.
\newblock {\em Soc. Forces} 70(1):19--42.

\bibitem{Chen2018Flexible}
Chen S, Mira A, Onnela JP (2018) Flexible model selection for mechanistic
  network models via super learner.
\newblock {\em arXiv:1804.00237 [stat.ME]}.

\bibitem{Lintusaari2017Fundamentals}
Lintusaari J, Gutmann MU, Dutta R, Kaski S, Corander J (2017) Fundamentals and
  recent developments in approximate bayesian computation.
\newblock {\em Syst. Biol.} 66(1):e66--e82.

\bibitem{Foster2011Clustering}
Foster DV, Foster JG, Grassberger P, Paczuski M (2011) Clustering drives
  assortativity and community structure in ensembles of networks.
\newblock {\em Phys. Rev. E} 84(6):066117.

\bibitem{Peng2015Assortative}
Peng TQ (2015) Assortative mixing, preferential attachment, and triadic
  closure: A longitudinal study of tie-generative mechanisms in journal
  citation networks.
\newblock {\em J. Informetr.} 9(2):250--262.

\bibitem{Mele2017Structural2}
Mele A (September 3, 2017) A structural model of homophily and clustering in
  social networks.
\newblock Available at SSRN: \url{https://ssrn.com/abstract=3031489} or
  \url{http://dx.doi.org/10.2139/ssrn.3031489}.

\bibitem{Mele2017Structural1}
Mele A (2017) A structural model of dense network formation.
\newblock {\em Econometrica} 85(3):825--850.

\bibitem{Snijders2006New}
Snijders TA, Pattison PE, Robins GL, Handcock MS (2006) New specifications for
  exponential random graph models.
\newblock {\em Sociol. Methodol.} 36(1):99--153.

\bibitem{Gross2008Adaptive}
Gross T, Blasius B (2008) Adaptive coevolutionary networks: A review.
\newblock {\em J. R. Soc. Interface} 5(20):259--271.

\bibitem{Gross2009Adaptive}
Gross T, Sayama H (2009) {\em Adaptive Networks: Theory, Models and
  Applications}.
\newblock (Springer).

\bibitem{Zschaler2012Adaptive}
Zschaler G (2012) Adaptive-network models of collective dynamics.
\newblock {\em Eur. Phys. J. Special Topics} 211(1):1--101.

\bibitem{Ito2001Spontaneous}
Ito J, Kaneko K (2001) Spontaneous structure formation in a network of chaotic
  units with variable connection strengths.
\newblock {\em Phys. Rev. Lett.} 88(2):028701.

\bibitem{Bornholdt2003Self}
Bornholdt S, R{\"o}hl T (2003) Self-organized critical neural networks.
\newblock {\em Phys. Rev. E} 67(6):066118.

\bibitem{Holme2006Nonequilibrium}
Holme P, Newman ME (2006) Nonequilibrium phase transition in the coevolution of
  networks and opinions.
\newblock {\em Phys. Rev. E} 74(5):056108.

\bibitem{Vazquez2007Time}
Vazquez F, Gonz{\'a}lez-Avella JC, Egu{\'\i}luz VM, San~Miguel M (2007)
  Time-scale competition leading to fragmentation and recombination transitions
  in the coevolution of network and states.
\newblock {\em Phys. Rev. E} 76(4):046120.

\bibitem{Iniguez2009Opinion}
I{\~n}iguez G, Kert{\'e}sz J, Kaski KK, Barrio RA (2009) Opinion and community
  formation in coevolving networks.
\newblock {\em Phys. Rev. E} 80(6):066119.

\bibitem{Axelrod1997Dissemination}
Axelrod R (1997) The dissemination of culture: A model with local convergence
  and global polarization.
\newblock {\em J. Conflict Resolut.} 41(2):203--226.

\bibitem{Davidsen2002Emergence}
Davidsen J, Ebel H, Bornholdt S (2002) Emergence of a small world from local
  interactions: Modeling acquaintance networks.
\newblock {\em Phys. Rev. Lett.} 88(12):128701.

\bibitem{Vazquez2003Growing}
V{\'a}zquez A (2003) Growing network with local rules: Preferential attachment,
  clustering hierarchy, and degree correlations.
\newblock {\em Phys. Rev. E} 67(5):056104.

\bibitem{Marsili2004Rise}
Marsili M, Vega-Redondo F, Slanina F (2004) The rise and fall of a networked
  society: A formal model.
\newblock {\em P. Natl. Acad. Sci. USA} 101(6):1439--1442.

\bibitem{Toivonen2006Model}
Toivonen R, Onnela JP, Saram{\"a}ki J, Hyv{\"o}nen J, Kaski K (2006) A model
  for social networks.
\newblock {\em Physica A} 371(2):851--860.

\bibitem{Timar2016Scale}
Tim{\'a}r G, Dorogovtsev SN, Mendes JFF (2016) Scale-free networks with
  exponent one.
\newblock {\em Phys. Rev. E} 94(2):022302.

\bibitem{Saramaki2014Persistence}
Saram{\"a}ki J, et~al. (2014) Persistence of social signatures in human
  communication.
\newblock {\em P. Natl. Acad. Sci. USA} 111(3):942--947.

\bibitem{Coleman1958Relational}
Coleman JS (1958) Relational analysis: the study of social organizations with
  survey methods.
\newblock {\em Hum. Organ.} 17(4):28--36.

\bibitem{Kivela2012Multiscale}
Kivel{\"a} M, et~al. (2012) Multiscale analysis of spreading in a large
  communication network.
\newblock {\em J. Stat. Mech.} 2012(03):P03005.

\end{thebibliography}

\clearpage
\begin{thebibliography}{1}

\bibitem{SI_Mathematica}
{Wolfram Research, Inc.} (2017) Mathematica, version 11.1.

\bibitem{SI_Onnela2007Mobile}
Onnela JP, et~al. (2007) Structure and tie strengths in mobile communication
  networks.
\newblock {\em Proceedings of the National Academy of Sciences}
  104(18):7332--7336.

\bibitem{SI_Adamic2005polblogs}
Adamic LA, Glance N (2005) The political blogosphere and the 2004 u.s.
  election: Divided they blog in {\em Proceedings of the 3rd International
  Workshop on Link Discovery}, LinkKDD '05.
\newblock (ACM, New York, NY, USA), pp. 36--43.

\bibitem{SI_Traud2012facebook100}
Traud AL, Mucha PJ, Porter MA (2012) Social structure of facebook networks.
\newblock {\em Physica A: Statistical Mechanics and its Applications}
  391(16):4165 -- 4180.

\bibitem{SI_Stehle2011Sociopatterns}
Stehlé J, et~al. (2011) High-resolution measurements of face-to-face contact
  patterns in a primary school.
\newblock {\em PLOS ONE} 6(8):e23176.

\end{thebibliography}
\end{document}